\documentclass[aip,jcp,preprint,footinbib,floatfix,10pt]{revtex4-1}
\usepackage[utf8]{inputenc}

\usepackage{amsmath}%
\usepackage{amssymb}%
\usepackage{graphicx}%
\usepackage{dcolumn}%
\usepackage{bm}%
\usepackage[caption=false]{subfig}%
\usepackage{multirow}%
\usepackage{verbatim}%
\usepackage{hyperref}%
\usepackage{xcolor}%

\begin{document}

\title{Coupling of state-resolved rovibrational coarse-grain model for nitrogen to stochastic particle method for simulating internal energy excitation and dissociation}

\date{\today}

\author{Erik \surname{Torres}}
\email[Corresponding author: ]{torres@vki.ac.be}
\affiliation{Aeronautics and Aerospace Department, von Karman Institute for Fluid Dynamics, Chauss\'ee de Waterloo 72, 1640 Rhode-Saint-Gen\`ese, Belgium}
\author{Thierry E. \surname{Magin}}
\affiliation{Aeronautics and Aerospace Department, von Karman Institute for Fluid Dynamics, Chauss\'ee de Waterloo 72, 1640 Rhode-Saint-Gen\`ese, Belgium}

\begin{abstract}
 We propose to couple a state-resolved rovibrational coarse-grain model to a stochastic particle method for simulating internal energy excitation and dissociation of a molecular gas. A coarse-grained model for a rovibrational reaction mechanism of an \emph{ab initio} database developed at NASA Ames Research Center for the $\mathrm{N_2}$-$\mathrm{N}$ system is modified based on variably-spaced energy bins. Thermodynamic properties of the modified  coarse-grained model allow us to closely match those obtained with the full set of rovibrational levels over a wide temperature range, while using a number of bins significantly smaller than the complete mechanism. The chemical-kinetic behavior of equally- and variably- spaced bin formulations is compared by simulating internal energy excitation and dissociation of nitrogen in an adiabatic, isochoric reactor. We find that the variably-spaced formulation is better suited for reproducing the dynamics of the full database at conditions of interest in Earth atmospheric entry. Also in this paper, we discuss details of our particle method implementation for the URVC bin model and describe changes to the Direct Simulation Monte Carlo (DSMC) collision algorithm, which become necessary to accommodate our state-resolved reaction mechanism for excitation and dissociation reactions. The DSMC code is then verified against equivalent master equation calculations. In these simulations, state-resolved cross sections are used in analytical form. These cross sections verify micro-reversibility relations for the rovibrational bins and allow for fast execution of the DSMC code. In our verification calculations, we obtain very close agreement for the concentrations profiles of $\mathrm{N}$ and $\mathrm{N_2}$, as well as the translational and rovibrational mode temperatures obtained independently through both methods. In addition to macroscopic moments, we compare discrete internal energy populations predicted at selected time steps via DSMC and the master equations. We observe good agreement between the two sets of results within the limits imposed by statistical scatter, which is inherent to particle-based DSMC solutions. As future work, the rovibrational coarse-grain model coupled to the particle method will allow us to study 3D  reentry flow configurations.
\end{abstract}


\maketitle


\section{Introduction} \label{sec:introduction}

Predicting the flow field around a space craft entering Earth's atmosphere at hypersonic speeds is a challenging task. At flight speeds typical of orbital return, a strong bow shock is formed ahead of blunt-nosed reentry vehicles and severe compression causes the gas temperature to increase thousands of kelvin above free-stream values. At such high temperatures, molecular species become vibrationally and electronically excited. They may also partially dissociate and ionize. A fraction of the newly formed atomic species recombine in the boundary layer near the vehicle's surface and release a significant amount of heat in the process. At the low-density conditions of the upper atmosphere, these processes occur at characteristic time scales of the same orders of magnitude as the hydrodynamic one, which means that for most of the reentry trajectory the gas surrounding the space craft exists in a state of thermochemical nonequilibrium. This can have a strong influence on flow field features, such as shock standoff distance and thickness, but also on the heat loads that the space craft has to endure. As the vehicle descends towards Earth it is exposed to free-stream gas with increasing density, from nearly zero at the outer edge of the atmosphere to the order of $10^{25}$ molecules per cubic meter at sea level. This increase in density is tied to an inversely proportional decrease of the gas' local mean free path. At higher altitudes higher than 60-70~km, rarefied gas effects play an important role, strongly affecting aerodynamic coefficients and, to a lesser degree, the heat flux on the vehicle's surface. The rarefied flow regime lies beyond the range of validity of the Navier-Stokes equations and one must resort to alternative simulation methods, such as particle-based Direct Simulation Monte Carlo (DSMC)~\cite{bird94a}, which solves the Boltzmann equation in a stochastic manner. At lower altitudes the flow around an Apollo-sized capsule can be regarded as fully within the continuum regime and is accurately described by classical Computational Fluid Dynamics (CFD) methods. 

Accurately modeling thermo-chemical nonequilibrium effects in reentry flows using CFD/DSMC methods requires reliable data for predicting the rates of internal energy excitation and molecular dissociation of the atmosphere's constituents. For many years, computational fluid dynamicists have relied on semi-empirical correlations based on the Landau-Teller model~\cite{landau36a, vincenti65a} for calculating the rates of vibrational relaxation. During the 1950's and 60's, experiments to estimate the characteristic times of vibrational relaxation for different diatomic molecules were carried out by Blackman~\cite{blackman56a}, Millikan and White~\cite{millikan63a} and Hanson and Baganoff~\cite{hanson70a}. Simultaneously, experiments were performed in shock-tube facilities to measure the dissociation rates of nitrogen by Byron~\cite{byron66a}, and  Appleton et al~\cite{appleton68a}. Hornung~\cite{hornung72a}, among others found evidence of a delay in the onset of dissociation downstream of the shock wave front, which he referred to as an induction time. From a theoretical viewpoint, this phenomenon is believed to be caused by ``vibrational favoring'' of the dissociation reaction~\cite{hammerling59a, treanor62a, marrone63a}. That is, $\mathrm{N_2}$-molecules populating higher-lying vibrational levels are much more likely to dissociate than those in the ground vibrational state, since they are already much closer to overcoming the reactions' energy barrier. However, in a cold gas these higher-lying vibrational levels must first become populated before any significant production of atomic nitrogen can occur. The induction time is thus believed to be tied to the time required for the initially cold free-stream nitrogen molecules behind the shock front become vibrationally excited. 

The desire to match the experimentally measured dissociation rates in CFD simulations led to the development of Park's two-temperature model~\cite{park89a, park90a}. In this model, dissociation rate coefficients are evaluated at a combined temperature $\sqrt{T T_\mathrm{v}}$ as a way to account for vibrational favoring. In Park's model, $T$ represents the temperature of the translational mode, and it is implicitly assumed that the rotational relaxation time is small enough for both modes remain in equilibrium throughout the shock. Meanwhile, the electronic and vibrational modes are assumed to follow a different Boltzmann distribution at $T_\mathrm{v}$. Despite well-known deficiencies~\cite{olejniczak95a, park10a}, Park's model is still the standard for describing thermochemical nonequilibrium effects in most production-level CFD codes for hypersonic applications~\cite{gnoffo89a, candler15c}, and the rate coefficients to fit his model are widely used in simulations of high-temperature reacting flows~\cite{park93a, park94a}. At orbital reentry speeds the rotational and vibrational modes become intimately mixed and the Park model does not apply~\cite{panesi13a, zhu16a}. However, even if one neglects this rovibrational coupling, there is enough evidence that nonequilibrium internal energy distributions cannot be adequately described by a single temperature. For instance, Davis~\cite{davis02a} has shown that nonequilibrium vibrational energy distributions are not well-captured by $T_\mathrm{v}$ when the departure from a Boltzmann distribution is significant. More recently, Guy et al.~\cite{guy13a, guy15a} have proposed a multi-internal temperature model to better capture the shape of the vibrational manifold, but in general the problem can only be overcome by resorting a full state-to-state approach~\cite{panesi13a}. Unfortunately, such rovibrational-state-resolved models are too computationally costly to be used in large-scale CFD simulations and several coarse-graining techniques, discussed later in this section, have been proposed to remedy the problem.

Meanwhile, the development of thermo-chemical nonequilibrium models for DSMC proceeded on a different path. Since DSMC is based on simulator particles representing the gas flow at the kinetic time scale, one might assume that detailed microscopic models describing the interactions of individual molecules during inelastic, or reactive collisions would naturally be incorporated into the method. However, this pre-supposes the availability of a full set of cross sections for all inelastic state transitions and chemical reactions relevant to the given gas mixture. The prohibitive computational cost of simulating detailed chemistry, both in terms of computer memory and central processing unit (CPU) time, coupled with the lack of reliable cross section data for practically all reactions of interest in high-temperature air chemistry meant that for many decades little progress could be made. It was not until Borgnakke and Larsen~\cite{borgnakke75a} proposed their phenomenological model for inelastic collisions that it became feasible -and more widespread- to include such effects in DSMC simulations. Their approach did not rely on detailed microscopic cross sections, but on simplified transition probabilities tuned to match characteristic times of rotational-translational and vibrational-translational relaxation ($\tau_\mathrm{RT}$, $\tau_\mathrm{VT}$) inferred from experiments, or predicted by theory~\cite{parker59a}. Although considerable effort has been invested in refining the Borgnakke-Larsen (BL) model such that the predicted rates of thermal relaxation match those obtained using Landau-Teller-type equations in CFD~\cite{lumpkin91a, haas94a, gimelshein02b, zhang13a}, discrepancies between DSMC, CFD and experimental results still exist, in particular at high reentry speeds~\cite{park10a}. The discrepancies between CFD and DSMC are partly due to there not existing a direct equivalent to the Bograkke-Larsen model at the hydrodynamic scale, but also because of as yet poorly understood rotation-vibration and dissociation coupling effects. One of the most widely used models for chemical reactions in DSMC was originally proposed by Bird~\cite{bird78a, bird94a}. His total collision energy (TCE) model became the standard in the community and is still in use in most production-level DSMC codes~\cite{dietrich95a, ivanov98a}. The TCE model incorporates information about the internal state of the colliding molecules by evaluating the reaction cross section at the sum of the collision pair's rotational, vibrational and relative translational energy. Post-reaction energy repartition is handled as in the BL-model. However, one of the drawbacks of the TCE model is that it fails to predict vibrational favoring in dissociation behind strong shock waves. Modifications to the TCE model, which include some form of vibrational favoring in the dissociation process have been proposed~\cite{haas93a, boyd93c}, but have not seen widespread adoption. 

Due to their relative ease of implementation, the combination of the phenomenological Borgnakke-Larsen and TCE models has remained the standard for high-temperature chemistry in DSMC for many years. Nevertheless, development of more physics-based thermo-chemistry models which incorporate molecular-scale information was not completely abandoned. As computers have became more powerful and the numerical tools of quantum chemistry have matured, several independent groups have made progress on introducing \emph{ab initio} reaction rate data into the DSMC method, in order to simulate hypersonic nonequilibrium flows. The Bari group was one of the first to implement a vibrationally-resolved reaction mechanism for nitrogen in a DSMC code~\cite{esposito99a, esposito00a, bruno02a}. In their work, they resolved 66 discrete vibrational levels of the $\mathrm{N_2}$-molecule. For diatom-atom collisions, i.e. $\mathrm{N_2} \left( v \right) + \mathrm{N}$, they used cross-section data based on quasi-classical trajectory (QCT) calculations. Another detailed set of cross-sections is based on the \emph{ab initio} QCT calculations of Jaffe et al.~\cite{jaffe08a} for the $\mathrm{N_2(X^1\Sigma_g^+)} + \mathrm{N({}^4S_u)}$-system, originally compiled at NASA Ames Research Center. This database has been used extensively in CFD studies of the dynamics of internal energy excitation and dissociation in nitrogen mixtures by Magin et al~\cite{magin10a, panesi13a}. Simultaneously, Kim and Boyd integrated the full set of Ames $\mathrm{N_2}$+$\mathrm{N}$ cross sections into a DSMC solver~\cite{kim14a}. In the present paper, we have used this same set of cross sections as the source for our reduced-size nonequilibrium model. More recently, Parsons et al~\cite{parsons14a} generated state-resolved dissociation cross sections for $\mathrm{N_2}$+$\mathrm{N_2}$-collisions also using the QCT method, but based on a different \emph{ab initio} potential energy surface (PES) for the $\mathrm{N_2(X^1\Sigma_g^+)}$+$\mathrm{N_2(X^1\Sigma_g^+)}$-system~\cite{paukku13a}.

All of the groups mentioned more or less followed the approach of using the QCT method on \emph{ab initio} potential energy surfaces to generate rovibrational-state-resolved inelastic collision- and dissociation cross sections, and/or corresponding rate coefficients. However, the amount of data obtained via this approach still remains too vast for efficient flow simulations in large-scale CFD, or DSMC codes. The fact that the $\mathrm{N_2}$-molecule possesses nearly 10000 rovibrational levels in just the ground electronic state means that the number of different possible rovibrational level transitions for $\mathrm{N_2}\left(v,J\right)$+$\mathrm{N} \rightarrow \mathrm{N_2}\left(v^\prime,J^\prime\right)$+$\mathrm{N}$ is on the order of $10^{8}$. The approach becomes completely intractable, if $\mathrm{N_2}\left(v_1,J_1\right)$+$\mathrm{N_2}\left(v_2,J_2\right)$ $\rightarrow$ $\mathrm{N_2}\left(v_1^\prime,J_1^\prime\right)$+$\mathrm{N_2}\left(v_2^\prime,J_2^\prime\right)$ are also taken into account, because the number of transitions potentially is of the order of $10^{16}$. 

As a way to circumvent this problem, two different approaches have been developed. On the one hand, the classical trajectory calculation DSMC (CTC-DSMC) method proposed by Koura~\cite{koura97a, koura98b, koura02a} replaces the stochastic collision models used in DSMC with actual trajectory calculations integrated on a PES. At the time, Koura's calculations were restricted to simulating rotational-translational nonequilibrium across shock waves at moderate Mach numbers, due to the lack of availability of high-fidelity PES's. But his approach inspired Schwartzentruber's group to develop their Direct Molecular Simulation (DMS) method~\cite{norman13a}. They extended the CTC-DSMC method to rotating, vibrating, and dissociating molecules using the \emph{ab-initio} PESs for nitrogen from the University of Minnesota~\cite{paukku13a}. Since the DMS method performs collisions “on-the-fly” within a DSMC simulation, it only computes those trajectories which are the most probable, instead of pre-computing all possible transition cross sections as in the state-to-state approach~\cite{valentini15a}. This is very advantageous when dealing with molecular systems containing many discrete internal energy states, such as the N4-system~\cite{macdonald18b}.

The second approach has been to develop so-called coarse-grain models~\cite{munafo12a, magin12a, munafo14c, munafo15a, zhu16a, sahai17a} to reduce the size of the state-resolved reaction mechanism. The details of the reduction differ for each model, but the basic concept is always to approximate the behavior of the full set of levels with a much smaller number of suitably defined internal energy groups (often called ``bins''), whose properties are weighted averages over the properties of the individual constituting levels. This lumping-together of internal energy states automatically leads to a significant reduction in the number of associated state-to-state reaction rates and greatly reduces the cost of simulations. Of course, an error with respect to the full set of levels is introduced by each type of reduction, since a-priori assumptions about the level populations within the bins have to be made. From our viewpoint, the jury is still out on which of these coarse-grain methods is the optimal choice for both CFD and DSMC applications. In our accompanying paper~\cite{torres18a}, we discuss the main differences between the existing models and their relative advantages and disadvantages. Here, we and only recall the main points relevant for the present work. It should be noted that coarse-graining is not a recent idea. In the context of reducing a state-resolved cross-section set for molecular dissociation and relaxation, it was already used more than 30 years ago by Haug et al.~\cite{haug87a, haug92a} In this paper, the authors compared the effectiveness of different coarse-graining strategies, e.g. rotationally- vs. vibrationally averaged, as well as energy-based binning for reproducing the nonequilibrium chemistry of hydrogen.

Before proceeding we should clarify that the discussion in this paper is focused on the mixture of a single diatomic species (in our case $\mathrm{N_2}$) and the corresponding atomic species produced by its dissociation (i.e. $\mathrm{N}$). The coarse-grain model discussed in Sec.~\ref{sec:thermodynamic_properties_bins} and the DSMC algorithm presented in Sec.~\ref{sec:bin_model_dsmc_integration} are both limited to this particular case. We are deliberately not considering the effect of additional diatomic and atomic species in high-temperature air (e.g. $\mathrm{O_2}, \mathrm{O}, \mathrm{NO}$, etc.),  because modeling the chemical-kinetic behavior of such a gas is greatly complicated by the simultaneous competing dissociation-recombination and exchange reactions between all mixture components~\cite{lim84a} and is beyond the scope of this paper.

We consider a mixture of molecular and atomic nitrogen, with both species in their ground electronic states. The 9390 rovibrational levels of the $\mathrm{N_2}$ molecule have been grouped together into a much smaller number of discrete internal energy states. The lumping-together of rovibrational levels is done using the energy-based, uniform rovibrational collisional (URVC) bin model. As a result, our mixture is composed of $\mathrm{N_2} \left( k \right)$-molecules, where the bins with index $k \in \mathcal{K}_\mathrm{BP}$ encompass all bound- and pre-dissociated levels, plus atomic nitrogen $\mathrm{N}$. Applying the URVC binning approach~\cite{magin12a}, the level-specific reaction rate/cross-section data from the Ames database were condensed into bin-resolved rate coefficients/cross sections for inelastic and reactive collisions between molecular and atomic nitrogen. More specifically, the bin-resolved reactions modeled in our current work are

\begin{enumerate}
 \item Excitation/deexcitation of $\mathrm{N_2} \left(k\right)$ by collision with N-atom to form $\mathrm{N_2} \left(l\right)$
 \item Dissociation of $\mathrm{N_2} \left(k\right)$ by collision with N-atom to form atomic nitrogen.
\end{enumerate}

In addition to regular inelastic collisions, so-called exchange reactions are considered, where one nitrogen atom is swapped between the two collision partners, as well as collisions for which the colliding molecule's pre- and post-collision internal energies remain unchanged (i.e. $\bar{E}_k = \bar{E}_l$).

The objective of this paper is two-fold. The first goal is to study the sensitivity of the gas' thermodynamic properties and chemical kinetics to the precise choice of URVC bins. For this purpose, in Sec.~\ref{sec:thermodynamic_properties_bins} we compare two distinct binning arrangements. The first one, already used when the URVC bin model was originally proposed~\cite{magin12a}, tends to better resolve the high-energy levels near the dissociation limit. Here we propose a new, alternative arrangement, which places more emphasis on resolving the lowest-energy bound levels. Through this change we better estimate the rovibrational energy content of cold free-stream gas in hypersonic flows. The thermodynamic properties of the gas mixture using a variety of bin arrangements are compared to those obtained with the full set of rovibrational levels. Then, in Sec.~\ref{sec:parametric_study_kinetics} the effect of our bin arrangements on the mixture's chemical kinetics is studied with the help of a simple benchmark test case. It consists of simulating internal energy excitation and dissociation of $\mathrm{N_2}$ in an adiabatic, constant-volume reactor. A range of initial conditions is used, in order to study the behavior of our updated bin model at different operating temperatures. For this study, we simulate the whole process by solving the associated system of master equations (see App.~\ref{sec:urvc_bin_model_macroscopic_balance_equations}), since this method is much faster than an equivalent DSMC solution, but provides equivalent information for our comparison. We show how the newly proposed bins do a better job at approximating the full set of levels than the original formulation.

The second goal of this paper is to show the details of our implementation of the URVC bin model in the DSMC method. In Sec.~\ref{sec:bin_model_dsmc_integration} we describe our changes to the DSMC algorithm, necessary for integrating the state-resolved cross sections for $\mathrm{N_2} \left( k \right)$+$\mathrm{N}$-collisions extracted from the Ames N3 database~\cite{torres18a}. Then, in Sec.~\ref{sec:verification} we verify our DSMC implementation against master equation results, using the same test case as in Sec.~\ref{sec:parametric_study_kinetics}. Although Sec.~\ref{sec:bin_model_dsmc_integration} is mainly concerned with comparing our DSMC implementation to the master equation results, in Secs.~\ref{sec:neglecting_recombination} and \ref{sec:dsmc_computational_cost} we discuss some of the limitations and the computational cost of our current DSMC implementation. Finally, in Sec.~\ref{sec:conclusions} we provide some conclusions on this work and propose future improvements.

We wish to show how we integrated the \emph{ab initio} data from the Ames N3 database into the DSMC method and intend to propose a pragmatic, energy-based binning approach as an alternative to more complex -and arguably computationally more expensive- binning strategies proposed recently. In particular, the two-dimensional rovibrational bins proposed by Zhu et al~\cite{zhu16a} require up to 100 discrete states to resolve the rovibrational energy levels. 


\section{Effect of URVC binning on macroscopic properties of \texorpdfstring{$\boldsymbol{\mathrm{N_2}}$}{N2-mixture}} \label{sec:thermodynamic_properties_bins} 

In this section, we discuss two different binning arrangements in more detail and show how the particular choice of bins affects the thermodynamic properties of the gas. For convenience's sake, the same 10-bin system used in our accompanying paper~\cite{torres18a} will be used as the starting point. 

\subsection{Original formulation: Equal-energy spacing for bound bins} \label{sec:equally_sized_bins}

To begin, recall that the full set of levels can be classified into bound and quasi-bound levels. Bound levels are those whose energy lies below $\Delta E_{\left( v=0,J=0 \right)}^D = 9.75 \, \mathrm{eV}$, while pre-dissociated levels all possess energies above this threshold. About 3/4 of all levels are bound and the remaining 1/4 are quasi-bound levels. Based on this ratio, one might naturally decide to assign 7 bins to the bound levels and use the remaining 3 bins to lump together the quasi-bound levels. As mentioned previously~\cite{magin12a}, care must be taken not to mix bound and pre-dissociated levels within the same bin. The initial choice was to use bins of constant ``energy spacing'' in the bound- and quasi-bound energy ranges respectively. For our 10-bin example the ``size'' of each bin containing bound levels is defined by a constant $\Delta E^B = \left( 9.75 \, \mathrm{eV} - 0 \, \mathrm{eV} \right) / 7 = 1.39 \, \mathrm{eV}$. For the pre-dissociated levels it is given by $\Delta E^P= \left( 14.92 \, \mathrm{eV} - 9.75 \, \mathrm{eV} \right) / 3 = 1.72 \, \mathrm{eV}$, where the upper limit corresponds to the energy of the highest-lying level. With these \emph{equally-spaced} intervals the sets of indices for the levels belonging to each bin, i.e. the $\mathcal{I}_k$'s, are automatically determined. 

Since it is known that the high-energy levels play an important role in the overall dynamics of $\mathrm{N_2}$-dissociation, it makes sense to spread out the relatively small number of pre-dissociated levels over the three highest-energy bins. The approach of using \emph{equally-spaced} energy intervals, with a fixed 7:3-ratio of bound- to pre-dissociated bins was used throughout all of the simulations using the URVC model, originally reported by Magin et al~\cite{magin12a}. However, their choice of bins is not unique, and in some situations alternative arrangements may be preferable~\cite{haug87a}. For example, one reason for shifting the emphasis away from the pre-dissociated levels and improving the bin model's resolution at the lower-energy limit could lie in the desire to more accurately describe the gas mixture's thermodynamic properties at lower temperatures. This applies to the free-stream gas encountered by space craft during entry into Earth's atmosphere, where typically $T_\infty \approx 300 \, \mathrm{K}$. At such low temperatures, only the lowest-lying rovibrational levels are populated to any significant degree. Any slight change in the distribution of $\mathrm{N_2}$-molecules among these levels would then imply a large shift in internal energy content of the gas mixture. 

A straightforward remedy could be to increase the overall number of bins until the difference in behavior of the binned system with respect to the full set of levels became negligible. In this way, fewer levels would be lumped together into a single bin, and the low-temperature resolution of the model would automatically improve. This approach is considered in the parametric study in Sec.~\ref{sec:parametric_study_thermo}, demonstrating that the equally-spaced bins can indeed deliver accurate results. However, using equally-spaced energy intervals requires more than 100 bins to reach satisfactory low-temperature accuracy. Since the original goal of introducing a coarse-grain model was to reduce the size of the detailed chemistry mechanism as much as possible, using such a large number of bins would in some sense defeat this purpose. Instead, the goal should be to re-arrange the existing 10 bins in such a manner as to better resolve the low-energy behavior of the whole set of levels.

\subsection{Alternative formulation: Variable-energy spacing for bound bins} \label{sec:variably_sized_bins}

In order to improve the behavior of the 10-bin system at lower temperatures, an alternative approach is proposed here. First, we re-distribute the existing 10 bins such that the first nine of them lie below the dissociation limit $\Delta E_{\left(v=0, J=0\right)}^D$. This results in an equal energy spacing of $\left( 9.75 \, \mathrm{eV} - 0 \, \mathrm{eV} \right) / 9 = 1.08 \, \mathrm{eV}$ for the 9 bins below the dissociation threshold. The pre-dissociated levels are all lumped together into a single bin in the energy interval $\left( 14.92 \, \mathrm{eV} - 9.75 \, \mathrm{eV} \right) / 1 = 5.17 \, \mathrm{eV}$. Second, re-define the energy limits between the first nine bins, such that their normalized size grows with the power $n$:
\begin{equation}
 \mathcal{E}_{k}^V = \Delta E_{\left( \substack{v=0,\\J=0} \right)}^D \left( \frac{\mathcal{E}_{k}^E}{\Delta E_{\left( \substack{v=0,\\J=0} \right)}^D} \right)^n, \quad k \in \mathcal{K}_\mathrm{B}. \label{eq:bin_energy_grid_transformation}
\end{equation}

\begin{figure}
 \centering
 \includegraphics[width=0.9\columnwidth]{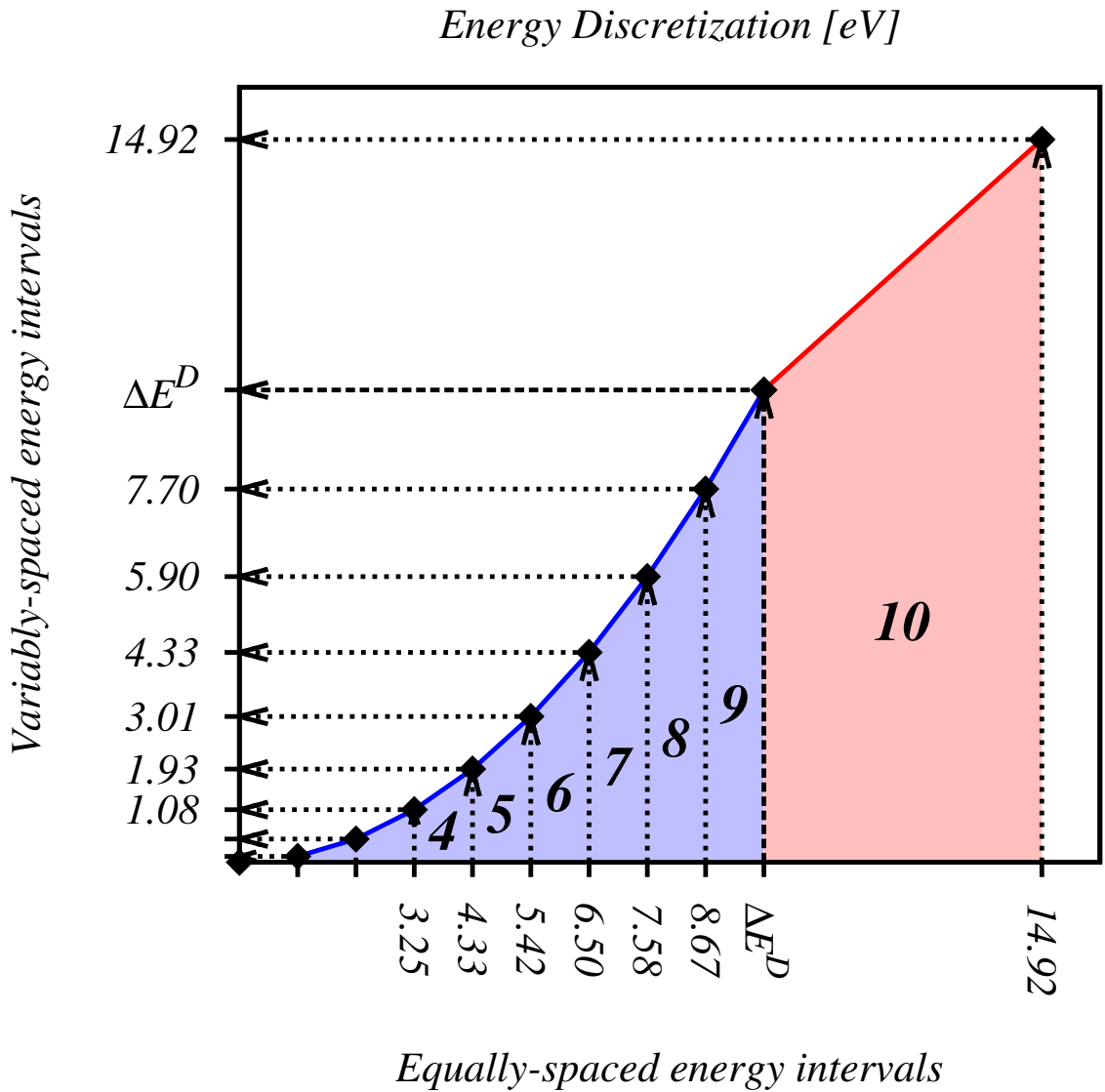}
 \caption{Re-scaling of original equally-spaced energy intervals $\Delta E_k^E$ (abscissa) to variably-spaced energy intervals $\Delta E_k^V$ (ordinate).}
 \label{fig:9plus1_bins_mesh}
\end{figure}

Here, the upper indices ``$E$'' and ``$V$'' stand for ``equally-spaced'' and ``variably-spaced'' energy intervals. The lower- and upper energy bounds of bin $k$ are labeled as $\mathcal{E}_k$ and $\mathcal{E}_{k+1}$ respectively, and the set $\mathcal{K}_\mathrm{B}$ includes all bins containing bound levels. The transformation from equally-spaced to the stretched bin boundaries is shown in Fig.~\ref{fig:9plus1_bins_mesh}, where an exponent $n = 2$ has been used. It becomes immediately clear that the width of the first few bins is much smaller than that in the original, equally-spaced approach.

After re-defining the bin boundaries in this manner, the updated bin-average quantities have been listed in Table~\ref{tab:9plus1_bins_variable_size_numbers}. Notice that the first bin now only contains 22 rovibrational levels. In fact, with this new approach the first three bins together contain fewer levels than the first bin in the original scheme of Table~II of Torres et al~\cite{torres18a}.
\begin{table}
 \centering
 \caption{Average properties of the full set of levels lumped into 10 bins. The 9 lower bins are composed exclusively of bound levels, the tenth bin of all the quasi-bound levels. An exponent $n = 2$ has been used to re-scale the bounds of the bins containing bound levels.}
 \label{tab:9plus1_bins_variable_size_numbers}

 \begin{tabular}{c | c | c | c | c}
  $\boldsymbol{k}$ & $\boldsymbol{i \in \mathcal{I}_k}$ & $\boldsymbol{\bar{g}_k}$ & $\boldsymbol{\mathcal{E}_k, \mathcal{E}_{k+1}}$ \textbf{[eV]} & $\boldsymbol{\bar{E}_k}$ \textbf{[eV]} \\ \hline
  1 & 1 \ldots 22 & 2145 & 0.00 \ldots 0.12 & 0.06 \\
  2 & 23 \ldots 72 & 9987 & 0.12 \ldots 0.48 & 0.33 \\
  3 & 73 \ldots 201 & 36699 & 0.48 \ldots 1.08 & 0.82 \\
  4 & 202 \ldots 452 & 93810 & 1.08 \ldots 1.93 & 1.54 \\
  5 & 453 \ldots 889 & 201444 & 1.93 \ldots 3.01 & 2.51 \\
  6 & 890 \ldots 1574 & 388311 & 3.01 \ldots 4.34 & 3.72 \\
  7 & 1575 \ldots 2615 & 697080 & 4.34 \ldots 5.90 & 5.17 \\
  8 & 2616 \ldots 4227 & 1228542 & 5.90 \ldots 7.71 & 6.87 \\
  9 & 4228 \ldots 7421 & 2434401 & 7.71 \ldots 9.75 & 8.86 \\ \hline
 10 & 7422 \ldots 9390 & 3109059 & \, 9.75 \ldots 14.92 & 11.25 \\
 \end{tabular}
\end{table}

\begin{figure}
 \centering
 \includegraphics[width=\columnwidth]{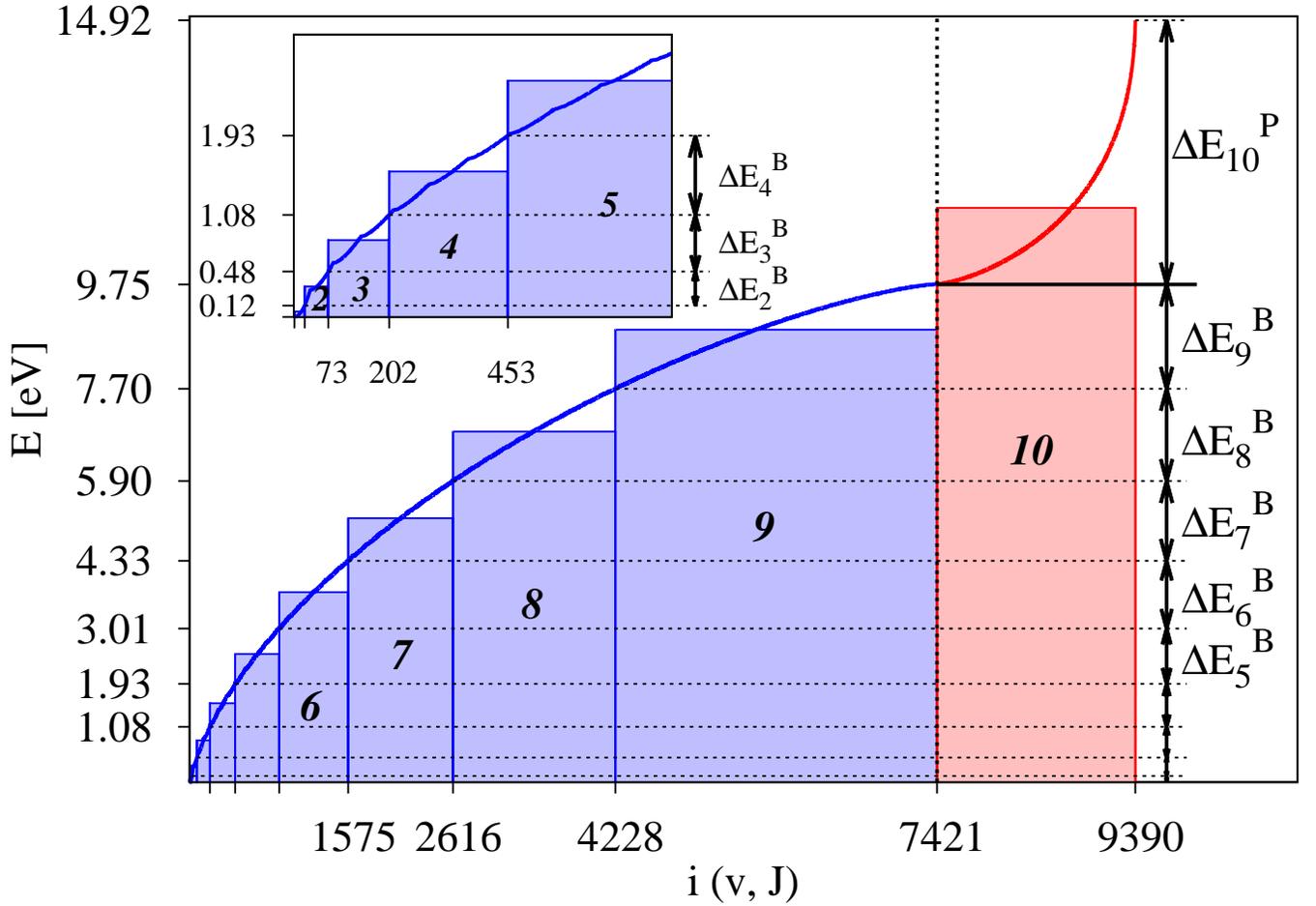}
 \caption{Altered 10-bin system with variable spacing of the energy intervals $\Delta E_k^B$ for the lower 9 bins, which contain only bound levels. The last bin contains all the pre-dissociated levels lying above $\Delta E _{\left(v=0, J=0\right)}^D = 9.75 \, \mathrm{eV}$. A close-up of the lowest-lying energy bins is shown as an insert.}
 \label{fig:9plus1_bins_variable_size_levels}
\end{figure}

The corresponding graphical representation of this new bin arrangement can be seen in Fig.~\ref{fig:9plus1_bins_variable_size_levels}. The first 4-5 bins are now too small to be clearly distinguished in the overall view, and are shown separately in a close-up window. Along with the newly defined ``heights'' of each bin, their ``width'' also grows more quickly with increasing energy, due to an increasing number of levels being lumped together as the bin number increases. 

In what follows we will refer to the two bin arrangements discussed here as \emph{equally-spaced} and \emph{variably-spaced} bins respectively. Bin ratios of bound-to-predissociated of 7:3 and 9:1 respectively were used as templates for all combinations tested in Sec.~\ref{sec:parametric_study_thermo} and \ref{sec:parametric_study_kinetics}. Although we studied systems with overall bin numbers much greater than 10, our ultimate goal was to find a bin arrangement, which would closely match the behavior of the full set of levels, while using as small a number of bins as possible.

\subsection{Dependence of thermodynamic properties on bin arrangement} \label{sec:parametric_study_thermo}

We now examine the effect of two alternative bin arrangements on the thermodynamic properties of $\mathrm{N_2}$. For this comparison we varied the overall number of bins between 10 and 200. All tested combinations are listed in Table~\ref{tab:tested_bin_ratios}.

\begin{table}
 \centering
 \caption{Thermodynamic properties of URVC bin model: Tested combinations of truly-bound to pre-dissociated (B:P) bins}
 \label{tab:tested_bin_ratios}
 
 \begin{tabular}{c | c | c}
  Total & \emph{Equally-spaced}  & \emph{Variably-spaced} \\
  bins & ($n = 1$) & ($n = 2$) \\ \hline
  10 & 7:3 & 9:1 \\
  20 & 14:6 & 18:2 \\
  100 & 70:30 & 90:10 \\
  200 & 140:60 & 180:20
 \end{tabular}
\end{table}


The ``quality'' of each choice of bins is now judged by how well they approximate the specific thermal energy of pure $\mathrm{N_2}$:
\begin{equation}
 e_\mathrm{N_2} = \frac{3}{2} \frac{\mathrm{k_B} T}{m_\mathrm{N_2}} + \frac{1}{m_\mathrm{N_2}} \frac{\bar{A}_\mathrm{N_2}^\mathrm{int}}{\bar{Q}_\mathrm{N_2}^\mathrm{int}}, \label{eq:molecular_nitrogen_specific_internal_energy}
\end{equation}
and its corresponding constant-volume heat capacity:
\begin{equation}
 c_\mathrm{v, N_2} = \frac{3}{2} \frac{\mathrm{k_B}}{m_\mathrm{N_2}} + \frac{1}{m_\mathrm{N_2} \mathrm{k_B} T^2} \left[ \frac{\bar{B}_\mathrm{N_2}^\mathrm{int}}{\bar{Q}_\mathrm{N_2}^\mathrm{int}} - \left( \frac{\bar{A}_\mathrm{N_2}^\mathrm{int}}{\bar{Q}_\mathrm{N_2}^\mathrm{int}} \right)^2 \right], \label{eq:molecular_nitrogen_specific_cv}
\end{equation}
when compared to using the full set of levels. Both expressions depend on the partition function of the internal energy mode of $\mathrm{N_2}$, i.e. $\bar{Q}_\mathrm{N_2}^\mathrm{int} = \sum_{k \in \mathcal{K}_\mathrm{BP}} \left\lbrace \bar{g}_k \, \exp \left( - \bar{E}_k/ \mathrm{k_B} \, T \right) \right\rbrace$, as well as the terms $\bar{A}_\mathrm{N_2}^\mathrm{int} = \sum_{k \in \mathcal{K}_\mathrm{BP}} \left\lbrace \bar{g}_k \bar{E}_k \exp \left( - \bar{E}_k / \mathrm{k_B} T \right) \right\rbrace$ and $\bar{B}_\mathrm{N_2}^\mathrm{int} = \sum_{k \in \mathcal{K}_\mathrm{BP}} \left\lbrace \bar{g}_k \bar{E}_k^2 \exp \left( - \bar{E}_k / \mathrm{k_B} T \right) \right\rbrace$.

\begin{figure}
 \centering
 \includegraphics[width=\columnwidth]{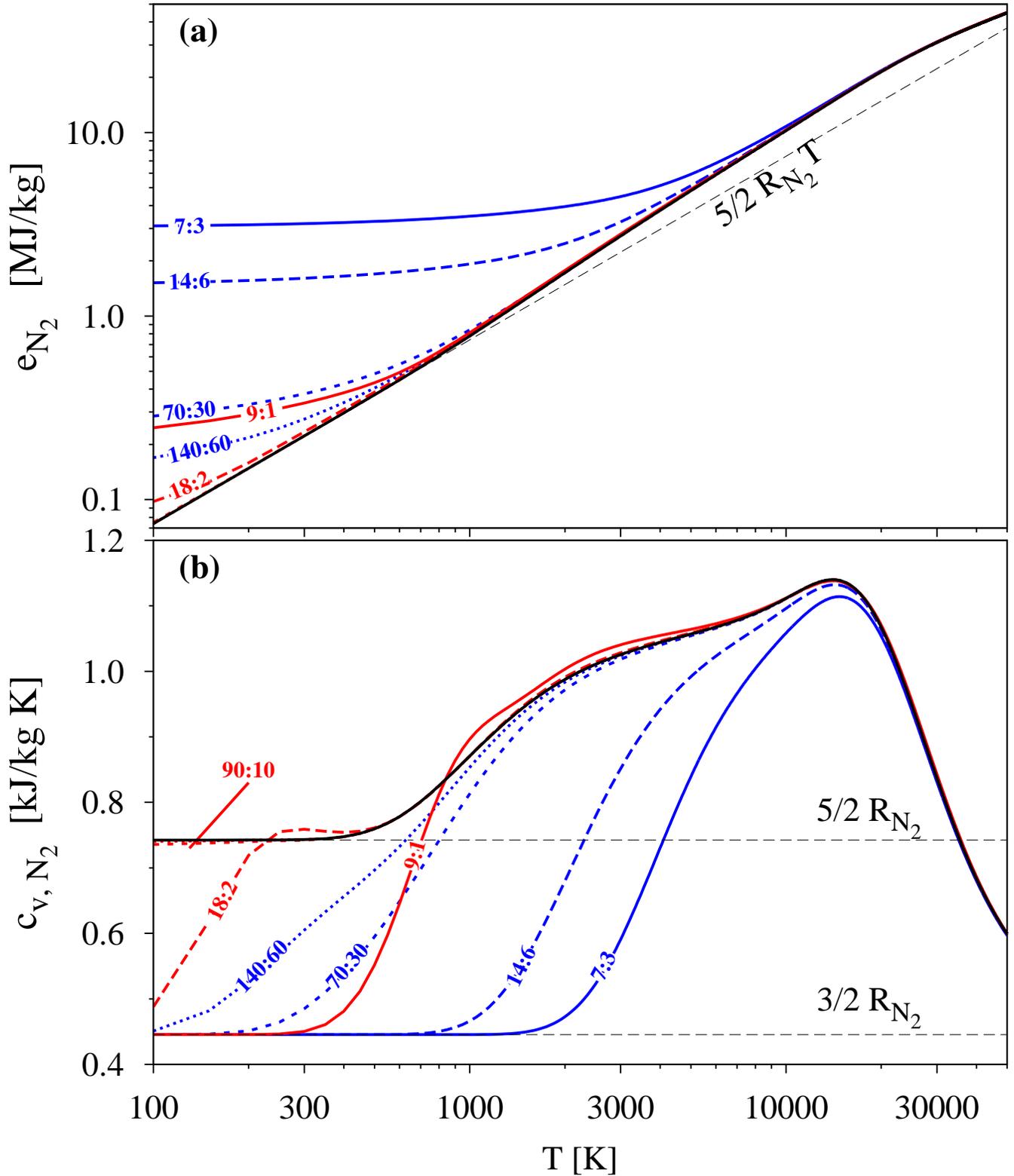}
 
 \caption{Thermodynamic properties of pure $\mathrm{N_2}$. Plot (a): specific thermal energy [$\mathrm{MJ / kg}$], plot (b): constant-volume heat capacity [$\mathrm{kJ / kg \, K}$]). Blue lines: URVC bin model with \emph{equally-spaced} bins ($n = 1$), Red lines: URVC bin model with \emph{variably-spaced} bins ($n = 2$). Black solid lines: reference properties with full set of rovibrational levels}
 \label{fig:thermodynamic_properties_1p0_vs_2p0}
\end{figure}

The comparisons are shown in Figs.~\ref{fig:thermodynamic_properties_1p0_vs_2p0}~(a) and (b) respectively, where the curves of $e_\mathrm{N_2}$ and $c_{v, \mathrm{N_2}}$ are plotted over the temperature range $100 \le T \le 50000 \, \mathrm{K}$. In order to better visualize the low-temperature region, logarithmic scales are used for both axes. Each set of bins is identified by a particular line color and pattern. Additionally, in Fig.~\ref{fig:thermodynamic_properties_1p0_vs_2p0}~(a) a solid black curve represents the thermal energy obtained with the full set of rovibrational levels. For further reference, a dashed black line labeled $\frac{5}{2} R_\mathrm{N_2} T$ represents the fraction of $e_\mathrm{N_2}$ corresponding to the sum of the fully excited translational and rotational modes of $\mathrm{N_2}$. Fig.~\ref{fig:thermodynamic_properties_1p0_vs_2p0}~(b) shows the corresponding constant-volume specific heat capacity of $\mathrm{N_2}$ for the same bin arrangements. As before, the solid black line represents the reference value obtained with the full set of levels. The two additional horizontal dashed lines, drawn at $\frac{3}{2} R_\mathrm{N_2}$ and $\frac{5}{2} R_\mathrm{N_2}$ represent the contributions to $c_{v, \mathrm{N_2}}$ of the translational- and the fully excited rotational mode respectively. Notice from the behavior of the solid black line that the heat capacity predicted by the full set of levels tends towards its limiting value of $\frac{5}{2} R_\mathrm{N_2}$ at temperatures below $\approx 400 \, \mathrm{K}$. Thus, the full set of levels correctly predicts that at these low temperatures the vibrational mode is practically not excited.

The profiles in Fig.~\ref{fig:thermodynamic_properties_1p0_vs_2p0} rather unsurprisingly confirm that the more overall bins are used, the better the agreement with the full set of levels becomes. The biggest mismatch with respect to the reference profile is observed at the lower temperature end, when using the equally-spaced bins (blue curves). In particular, the 7:3- and 14:6-bin cases over-predict $e_\mathrm{N_2}$ by almost two orders of magnitude in this limit. At the same time, the heat capacity is noticeably under-predicted for temperatures below 10000~K. As the overall number of equally-spaced bins is increased, the agreement with the full set of levels progressively improves. However, even with as many as 140:60 bins, significant errors can still be observed at temperatures around 300~K. In all cases the thermal energy tends to be over-estimated, while the heat capacity tends to be under-estimated. 

By contrast, the variably-spaced bins with a 9:1-ratio of bound-to-predissociated bins (red curves) follow the full set of levels much more closely. In fact,  for the 9:1- and 18:2-bin cases, deviations from the reference curves only become significant at temperatures below 1000~K. Any such differences almost completely disappear with the 90:10-bin system, where minute deviations from the reference curves can only be seen at temperatures below 300~K (see Fig.~\ref{fig:thermodynamic_properties_1p0_vs_2p0}~(b)). The curves for the 180:20-bin system have not been labeled at all, because they essentially lie on top of the reference curves over the whole temperature range. In summary, it is clear that for the same overall number of bins, the thermodynamic properties predicted with the 9:1-bin ratio agree much more closely with the full set of levels than when the original 7:3-bin ratio is used. Clearly, increasing the resolution of the bin model in the lower-energy range of the truly-bound levels is more beneficial than doing so in the higher-energy range of the pre-dissociated levels. This is to be expected, given that Eqs.~(\ref{eq:molecular_nitrogen_specific_internal_energy}) and (\ref{eq:molecular_nitrogen_specific_cv}) were computed assuming a Boltzmann distribution of the internal energy level populations. The thermodynamic properties are expected to be most sensitive to the populations of the lowest-energy levels, but much less so to barely populated high-energy tails.


\subsection{Influence of binning strategy on rovibrational relaxation and dissociation rates} \label{sec:parametric_study_kinetics}

So far, only the influence of the different bin arrangements on the thermodynamic properties has been examined. To study the effect of our model reduction on the chemical-kinetic behavior of the N3-system, one might want to compare the bin-resolved rate coefficients, or respective cross sections. However, due to the large number of reactions involved in even a small 10-bin system, we prefer to study the overall effect as part of a test case. 

We have chosen to simulate internal energy excitation and dissociation of an $\mathrm{N_2}$,$\mathrm{N}$-mixture in a constant-volume, adiabatic reactor. For simplicity, we have assumed spatially homogeneous gas properties, neglecting any flow field gradients and boundary effects. Thus, the problem is no longer dependent on spatial coordinates, and any variations in the system occur only as a function of time. Total mass, momentum and energy of the system are conserved, allowing us to reduce the governing equations of gas dynamics to a set of ordinary differential equations, which track the concentration of all mixture components over time, in addition to a conservation equation for the total energy. The computer code we use to solve this set of master equations has been used extensively in previous work~\cite{magin12a, panesi13a, munafo14c}. It is optimized for execution speed and is ideal for the parametric study presented in this section.

The initial conditions are summarized in Table~\ref{tab:case_2_reservoir_conditions}. They were chosen to approximate the extreme conditions in nitrogen gas, as they would be encountered immediately downstream of a strong shock wave. At $t = 0$, the gas is initialized in a state of thermal and chemical non-equilibrium, with the mixture translational temperature $T_\mathrm{t}^0 > T_\mathrm{int}^0$, while the initial bin populations of $\mathrm{N_2}$ follow a Boltzmann distribution at the initial internal temperature $T_\mathrm{int}^0 = 300$ K. A small amount of atomic nitrogen (arbitrary mass fraction $y_\mathrm{N}^0 = 0.014$) is added to the initial mixture. Including it is necessary to trigger the inelastic collisions and dissociation reactions, since our reaction mechanism derived from the Ames N3 database only includes detailed reaction rates for $\mathrm{N_2} \left( k \right)$+$\mathrm{N}$-collisions. This automatically implies that neither excitation/deexcitation, nor dissociation reactions for $\mathrm{N_2} \left( k \right)$+$\mathrm{N_2} \left( l \right)$ encounters are being modeled. The missing $\mathrm{N_2}$-$\mathrm{N_2}$ reaction rates have slightly different implications for the hydrodynamic-level master equation simulations presented in this section, as opposed to the kinetic-level DSMC simulations reported in Sec.~\ref{sec:dsmc_verification_test_case}. At the hydrodynamic level, neglecting inelastic and reactive $\mathrm{N_2} \left( k \right)$+$\mathrm{N_2} \left( l \right)$-collisions simply meant that the respective rate coefficients were assumed to be zero, and would therefore not influence the mass production rates of $\mathrm{N_2} \left( k \right)$ and $\mathrm{N}$. At the kinetic scale on the other hand, a workaround had to be implemented whenever the DSMC algorithm would randomly select two nitrogen molecules for collision. Since no reaction cross sections were available, it was assumed that any such $\mathrm{N_2} \left( k \right)$+$\mathrm{N_2} \left( l \right)$-collisions would simply result in isotropic scattering, without internal energy exchange. As discussed in Sec.~\ref{sec:verification}, the standard variable hard sphere (VHS) model~\cite{bird80a} was used to estimate these elastic cross sections.

We refer to the three sets of initial conditions in Table~\ref{tab:case_2_reservoir_conditions} as the low, medium and high-temperature (or velocity) cases. They differ only by the initial values of translational temperature and static pressure. These, in turn, were determined as the conditions downstream of a normal shock wave by applying the Rankine-Hugoniot jump relations. See App.~\ref{sec:initial_conditions_explanation} for the rationale behind selecting these conditions. The three sets of conditions selected serve to examine the behavior of the two binning arrangements when used in different temperature ranges and to test the ``robustness'' of our formulation. The same test case was repeated for both bin arrangements using 10, 100 and 1000 overall bins. As discussed in Sec.~\ref{sec:parametric_study_thermo}, increasing the overall number of bins effectively improves the ``resolution'' of the rovibrational levels in the bin model and brings the bins' thermodynamic properties closer to those of the full set of levels. While in Sec.~\ref{sec:parametric_study_thermo} it was observed that we are able to converge to the reference thermodynamic properties over the whole temperature range by using at least 90 bound and 10 quasi-bound variably-spaced bins, we should not automatically assume that the same applies to the chemical-kinetic properties. This is why in this section we are performing additional simulations using systems with up to 1000 bins. For the three different initial conditions we also performed reference simulations using the complete RVC model with all 9390 levels. 

Based solely on thermodynamic considerations and the imposed initial conditions, we can determine the final equilibrium temperatures and composition within the reservoir. These are listed in Table~\ref{tab:case_2_reservoir_conditions} for the three sets of initial conditions. For the batch of simulations using the original binning strategy, column 1 lists the number of bound- and predissociated bins actually used, followed by final equilibrium temperature, pressure and mass fraction of atomic nitrogen. The corresponding information for the alternative binning strategy is listed in columns 5-8 of Table~\ref{tab:case_2_reservoir_conditions}. Notice that for both of the largest binned systems the sum of bound- and predissociated bins does not exactly add up to 1000. This is because some of the relatively small energy intervals defined for these bins actually don't containing any levels. Thus, these ``empty'' bins are excluded from the overall count. The last row in all three sub-tables lists the corresponding quantities for the full RVC model. As expected, the more overall bins are used the better the agreement with these reference values becomes. Simultaneously, one can see that the alternative, variably-spaced binning strategy requires far fewer bins to converge to the reference values than the original equally-spaced bins.

\begin{table}
 \centering
 \caption{Internal energy excitation and dissociation in adiabatic reactor: Initial reservoir conditions at $t = 0$ and at final equilibrium states.} \label{tab:case_2_reservoir_conditions}

  \begin{tabular}{r c r c r c}
  \hline \hline
  \multicolumn{6}{c}{Initial nonequilibrium conditions} \\ \hline
                 & $u_\infty$[m/s] & $T_\mathrm{t}^0$ [K] & $T_\mathrm{int}^0$ [K] & $p_0$ [Pa] & $y_N^0$ \\ \hline
    Low temp.    & 6500 &  $28\,766$ & 300 & 453.9 & 0.014 \\
    Medium temp. & 9613 &  $62\,546$ & 300 & 3164.0 & 0.014 \\
    High temp.   & 13000 & $114\,160$ & 300 & 14241.0  & 0.014
  \end{tabular}

  \begin{tabular}{c c c c c c c c}
   \hline
   \multicolumn{8}{c}{Final equilibrium conditions} \\ \hline
   \multicolumn{4}{c}{Equally-sized, $n = 1$} & \multicolumn{4}{c}{Variably-sized, $n = 2$} \\ \hline
   \multicolumn{8}{c}{ Low temp. } \\ \hline
   B:P & $T$ [K] & $p$ [Pa] & $y_\mathrm{N}$ & B:P & $T$ [K] & $p$ [Pa] & $y_\mathrm{N}$ \\ \hline
   7:3 & 4756 & 98.78 & 0.335 & 9:1 & 4724 & 93.55 & 0.273 \\
   14:6 & 4769 & 96.77 & 0.304 & 18:2 & 4733 & 93.58 & 0.271 \\
   70:30 & 4743 & 94.09 & 0.275 & 90:10 & 4737 & 93.67 & 0.271 \\
   700:269 & 4737 & 93.67 & 0.271 & 739:98 & 4737 & 93.67 & 0.271 \\ \hline
   full & 4737 & 93.67 & 0.271 & & 4737 & 93.67 & 0.271 \\ \hline
   \multicolumn{8}{c}{ Medium temp. } \\ \hline
   B:P & $T$ [K] & $p$ [Pa] & $y_\mathrm{N}$ & B:P & $T$ [K] & $p$ [Pa] & $y_\mathrm{N}$ \\ \hline
   7:3 & 5763 & 512.3 & 0.782 & 9:1 & 5678 & 482.9 & 0.705 \\
   14:6 & 5737 & 497.2 & 0.737 & 18:2 & 5688 & 483.0 & 0.703 \\
   70:30 & 5699 & 485.2 & 0.707 & 90:10 & 5693 & 483.4 & 0.702 \\
   700:269 & 5693 & 483.4 & 0.703 & 739:98 & 5693 & 483.4 & 0.702 \\ \hline
   full & 5693 & 483.4 & 0.702 &        & 5693 & 483.4 & 0.702 \\ \hline
   \multicolumn{8}{c}{ High temp.} \\ \hline
   B:P & $T$ [K] & $p$ [Pa] & $y_\mathrm{N}$ & B:P & $T$ [K] & $p$ [Pa] & $y_\mathrm{N}$ \\ \hline
   7:3 & 24077 & 5923 & 1.00 & 9:1 & 20922 & 5146 & 1.00 \\
   14:6 & 22328 & 5493 & 1.00 & 18:2 & 20809 & 5119 & 1.00 \\
   70:30 & 20967 & 5158 & 1.00 & 90:10 & 20797 & 5116 & 1.00 \\
   700:269 & 20798 & 5116 & 1.00 & 739:98 & 20795 & 5116 & 1.00 \\ \hline
   full & 20797 & 5116 & 1.00 &        & 20797 & 5116 & 1.00 \\ \hline \hline
  \end{tabular}

\end{table}

\begin{figure}
 \centering
 \includegraphics[width=\columnwidth]{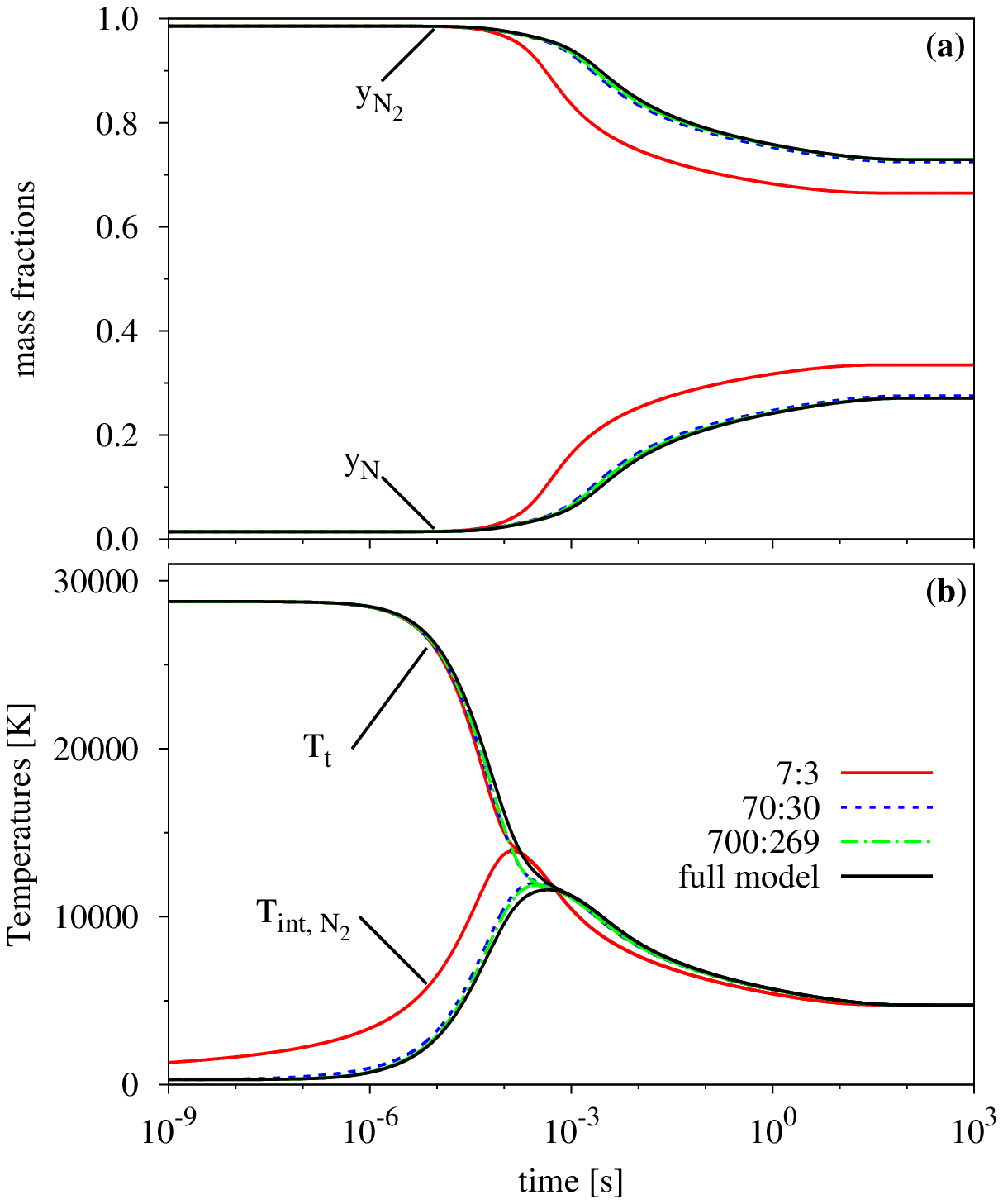}
 
 \caption{Excitation and dissociation of nitrogen in adiabatic reactor (\emph{Low-temp.} initial conditions, \emph{equally-spaced} bins). Red continuous lines: 10-bin (7:3), dashed blue lines: 100-bin (70:30), green dash-dotted lines: $\approx$1000-bin (700:269) system. Black continuous lines: reference solution with the full set of rovibrational levels. Plot (a): mass fractions of atomic and molecular nitrogen, plot (b): mixture translational temperature $T_\mathrm{t}$ and rovibrational mode temperature of molecular nitrogen $T_\mathrm{int, N_2}$.}
 \label{fig:master_equation_results_old_bins_lower_temp}
\end{figure}

\begin{figure}
 \centering
 \includegraphics[width=\columnwidth]{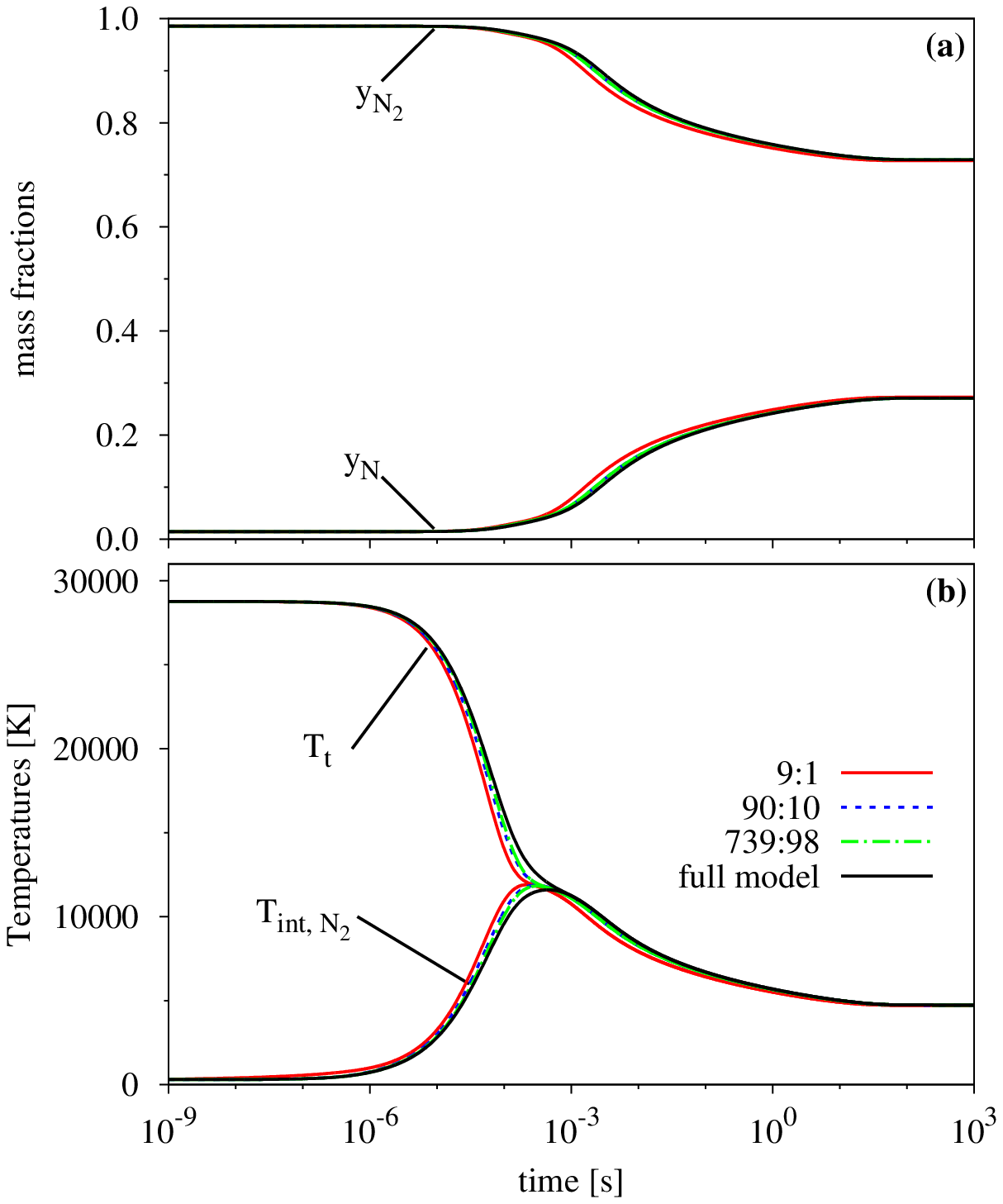}

 \caption{Excitation and dissociation of nitrogen in adiabatic reactor (\emph{Low-temp.} initial conditions, \emph{variably-spaced} bins). Red continuous lines: 10-bin (9:1), dashed blue lines: 100-bin (90:10), green dash-dotted lines: $\approx$1000-bin (739:98) system. Black continuous lines: reference solution with the full set of rovibrational levels. Plot (a): mass fractions of atomic and molecular nitrogen, plot (b): mixture translational temperature $T_\mathrm{t}$ and rovibrational mode temperature of molecular nitrogen $T_\mathrm{int, N_2}$.}
 \label{fig:master_equation_results_new_bins_lower_temp}
\end{figure}

\begin{figure}
 \centering
 \includegraphics[width=\columnwidth]{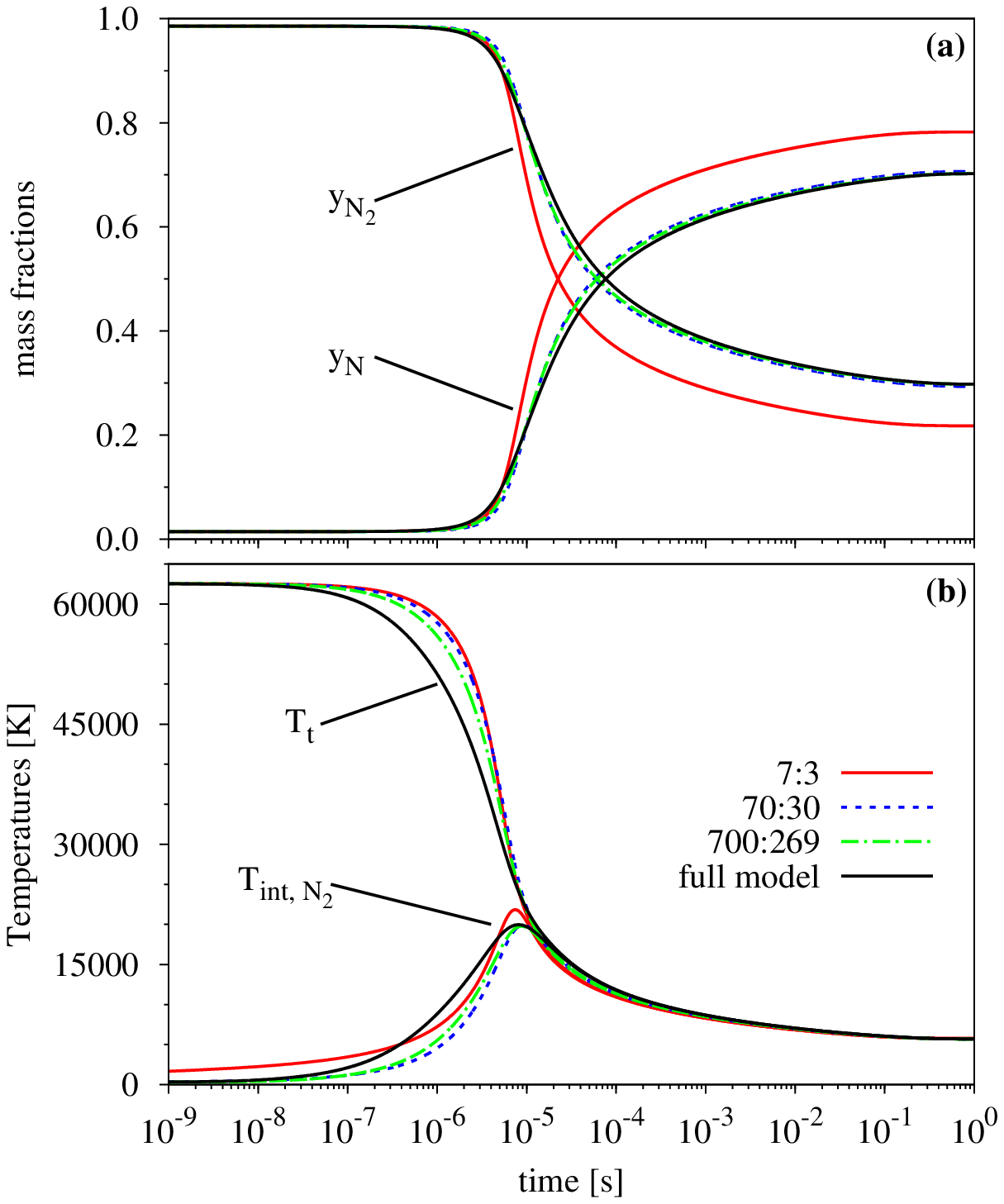}
 
 \caption{Excitation and dissociation of nitrogen in adiabatic reactor (\emph{Medium-temp.} initial conditions, \emph{equally-spaced} bins). Red continuous lines: 10-bin (7:3), dashed blue lines: 100-bin (70:30), green dash-dotted lines: $\approx$1000-bin (700:269) system. Black continuous lines: reference solution with the full set of rovibrational levels. Plot (a): mass fractions of atomic and molecular nitrogen, plot (b): mixture translational temperature $T_\mathrm{t}$ and rovibrational mode temperature of molecular nitrogen $T_\mathrm{int, N_2}$.}
 \label{fig:master_equation_results_old_bins}
\end{figure}

\begin{figure}
 \centering
 \includegraphics[width=\columnwidth]{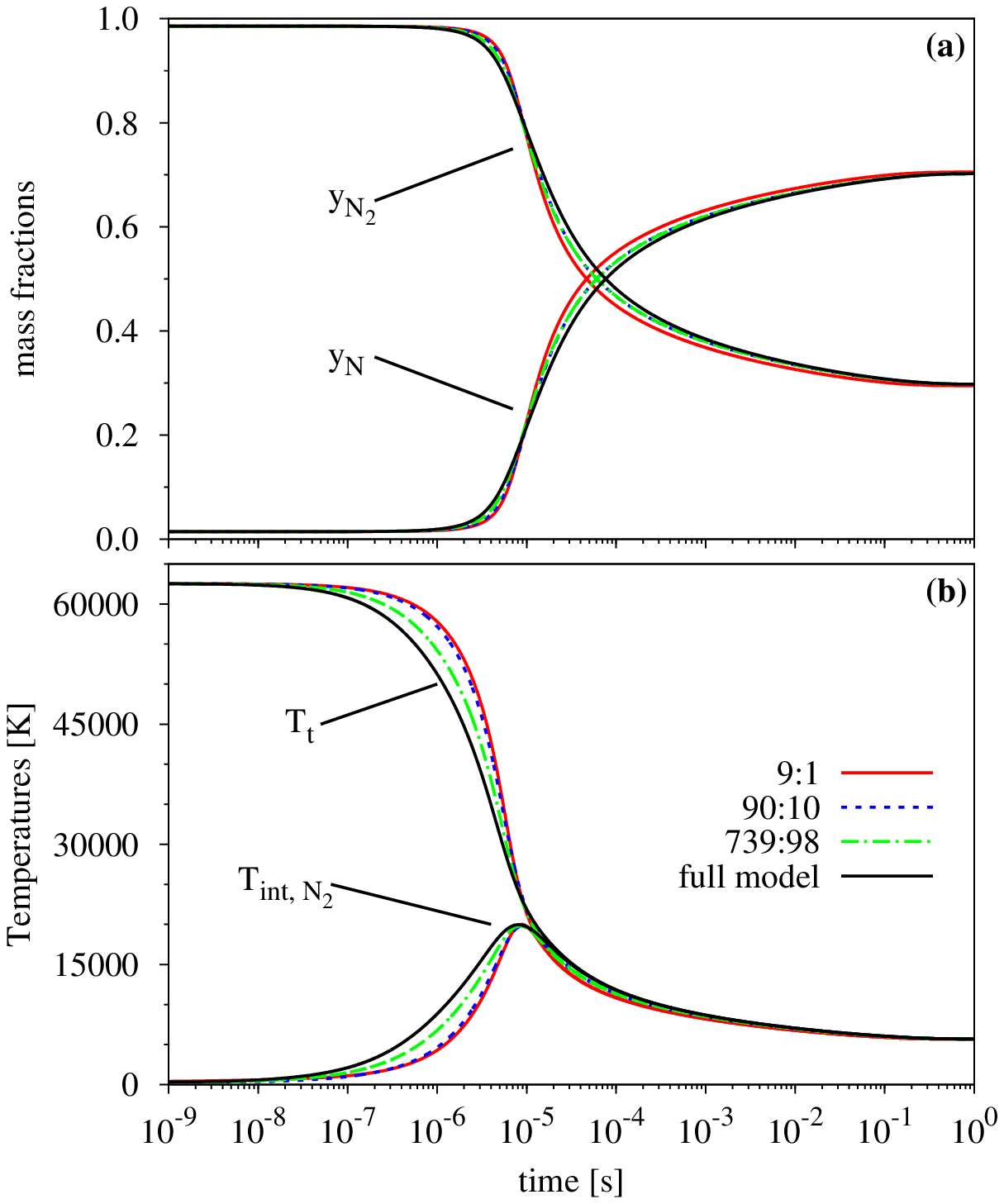}

 \caption{Excitation and dissociation of nitrogen in adiabatic reactor (\emph{Medium-temp.} initial conditions, \emph{variably-spaced} bins). Red continuous lines: 10-bin (9:1), dashed blue lines: 100-bin (90:10), green dash-dotted lines: $\approx$1000-bin (739:98) system. Black continuous lines: reference solution with the full set of rovibrational levels. Plot (a): mass fractions of atomic and molecular nitrogen, plot (b): mixture translational temperature $T_\mathrm{t}$ and rovibrational mode temperature of molecular nitrogen $T_\mathrm{int, N_2}$.}
 \label{fig:master_equation_results_new_bins}
\end{figure}

\begin{figure}
 \centering
 \includegraphics[width=\columnwidth]{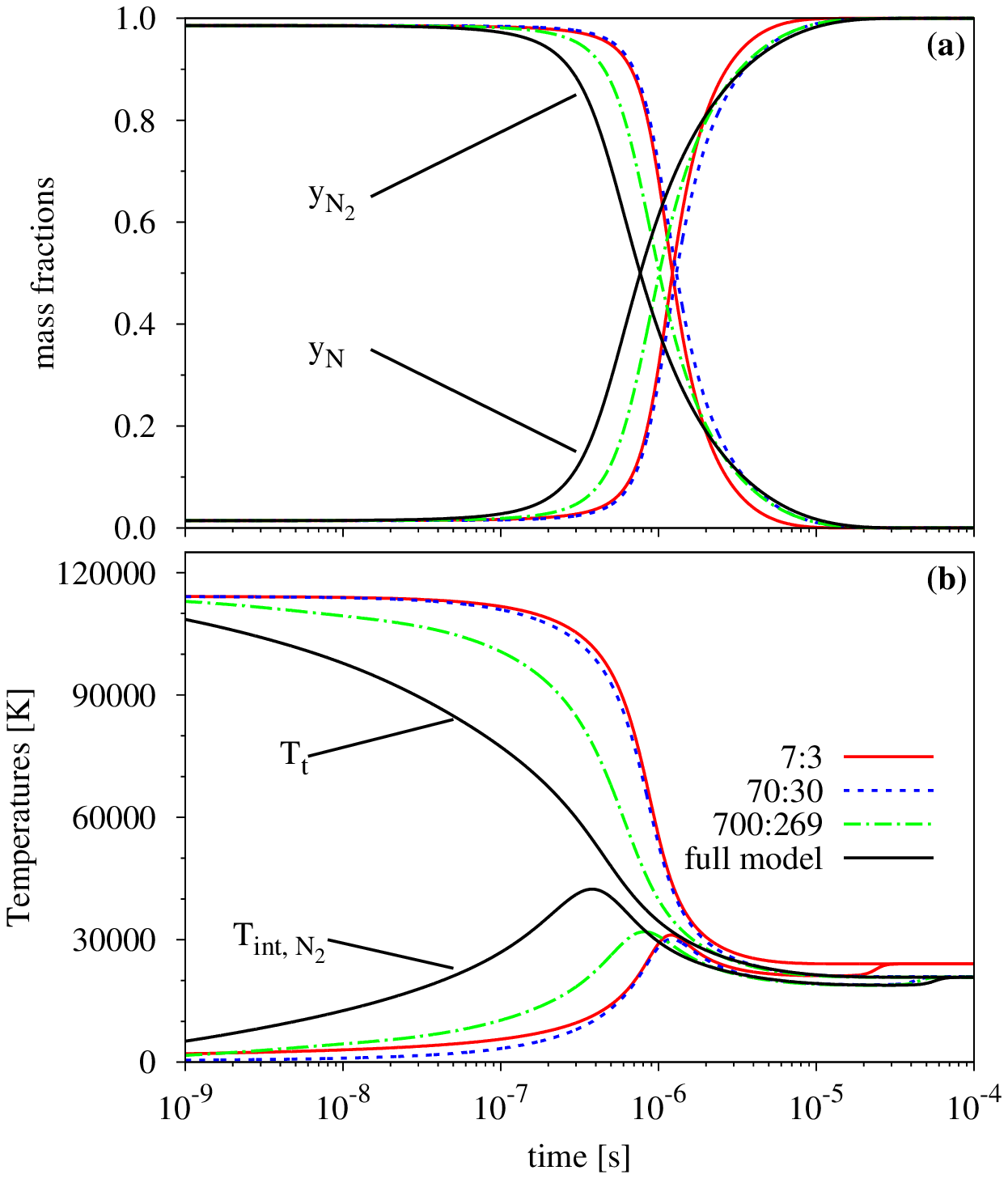}
 
 \caption{Excitation and dissociation of nitrogen in adiabatic reactor (\emph{High-temp.} initial conditions, \emph{equally-spaced} bins). Red continuous lines: 10-bin (7:3), dashed blue lines: 100-bin (70:30), green dash-dotted lines: $\approx$1000-bin (700:269) system. Black continuous lines: reference solution with the full set of rovibrational levels. Plot (a): mass fractions of atomic and molecular nitrogen, plot (b): mixture translational temperature $T_\mathrm{t}$ and rovibrational mode temperature of molecular nitrogen $T_\mathrm{int, N_2}$.}
 \label{fig:master_equation_results_old_bins_higher_temp}
\end{figure}

\begin{figure}
 \centering
 \includegraphics[width=\columnwidth]{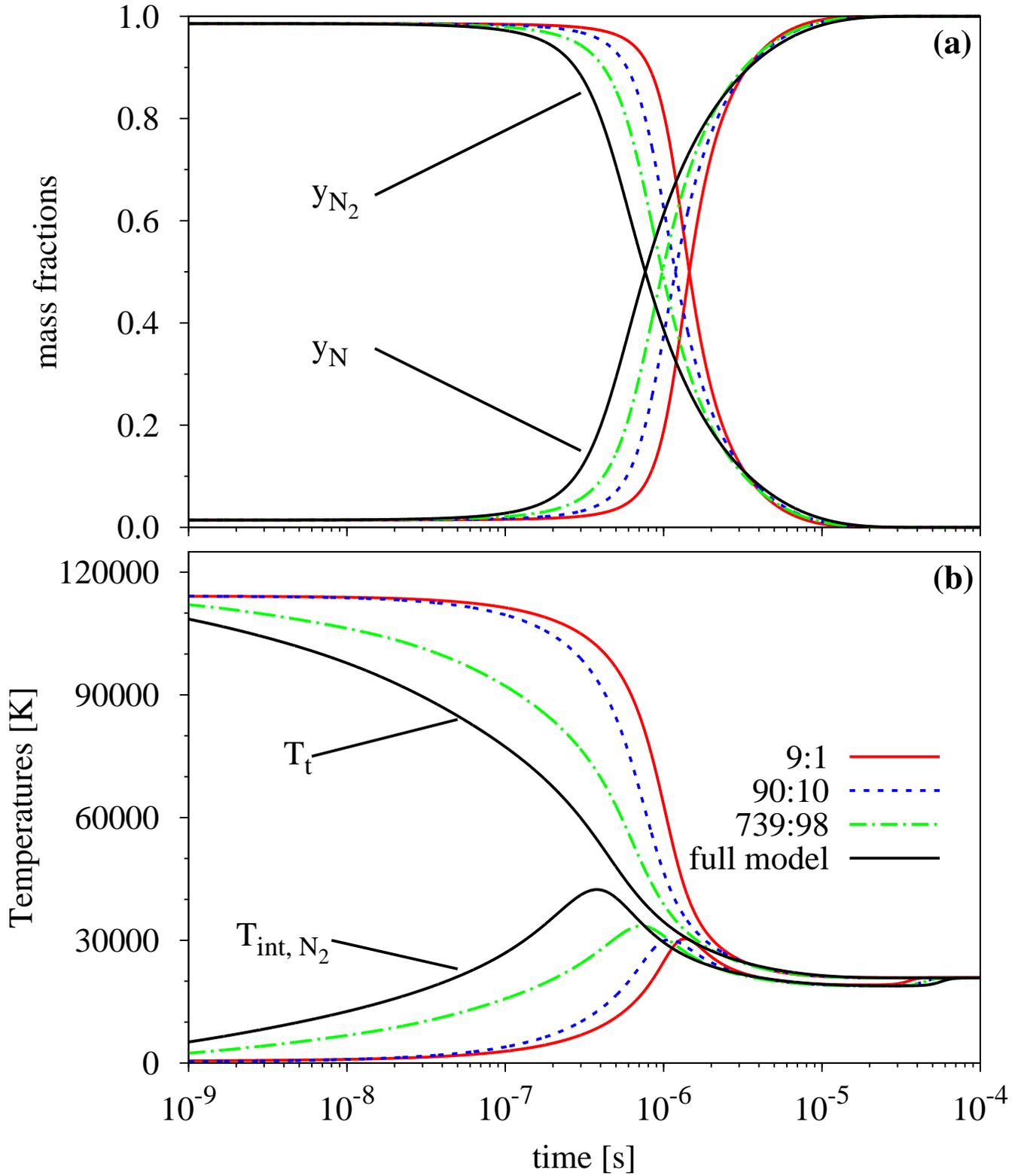}

 \caption{Excitation and dissociation of nitrogen in adiabatic reactor (\emph{High-temp.} initial conditions, \emph{variably-spaced} bins). Red continuous lines: 10-bin (9:1), dashed blue lines: 100-bin (90:10), green dash-dotted lines: $\approx$1000-bin (739:98) system. Black continuous lines: reference solution with the full set of rovibrational levels. Plot (a): mass fractions of atomic and molecular nitrogen, plot (b): mixture translational temperature $T_\mathrm{t}$ and rovibrational mode temperature of molecular nitrogen $T_\mathrm{int, N_2}$.}
 \label{fig:master_equation_results_new_bins_higher_temp}
\end{figure}

The time evolution of the mixture state in the reservoir for all combinations of bins and conditions is summarized in Figs.~\ref{fig:master_equation_results_old_bins_lower_temp} to \ref{fig:master_equation_results_new_bins_higher_temp}. Each figure shows the mass fractions of molecular- and atomic nitrogen (panel (a)), as well as the corresponding translational- and internal temperature profiles~\footnote{Refer to App.~\ref{sec:internal_temperature_definition} for the definition of $T_\mathrm{int}$ given nonequilibrium populations of rovibrational states.} (panel (b)). Due to the quick initial relaxation of translational and rovibrational modes, followed by a relatively slow approach to thermo-chemical equilibrium, a logarithmic scale has been used on the time axis. The general behavior in all conditions examined is roughly similar. At $t > 0$ the translational and internal modes, initially out of equilibrium, begin to relax. This causes the initial decrease in translational temperature, while the internal (rovibrational mode of $\mathrm{N_2}$) temperature begins to rise. Depending on the initial reservoir pressure and temperature (compare low- medium- and high-temperature cases), the moment at which both temperatures equalize, varies considerably. However, for all three initial states considered, the relaxation process seems to be well underway before dissociation becomes significant (as suggested by the delay in the production of atomic nitrogen). As more and more N-atoms are produced, the $\mathrm{N_2}$+ dissociation of $\mathrm{N_2}$ accelerates and begins to absorb a large part of the thermal energy in the gas. Therefore, the common reservoir temperature continually decreases from this point on, until slowly reaching its final equilibrium value.

Results for the low-temperature case using the original, equally-spaced bins are shown in Fig.~\ref{fig:master_equation_results_old_bins_lower_temp}, the corresponding results for the alternative, variably-spaced bins are plotted in Fig.~\ref{fig:master_equation_results_new_bins_lower_temp}. Since this case uses the lowest of the three initial translational temperatures, its final temperature of approximately 4737~K is also the lowest. As can be seen, at the given equilibrium temperature and pressure, a substantial fraction of the final gas remains as molecular nitrogen. Using the original formulation with 7:3 bins (solid red curves in Fig.~\ref{fig:master_equation_results_old_bins_lower_temp}), a noticeable difference in the time evolution of temperatures and species mass fractions is observed when compared to the reference profiles (solid black curves). While the relaxation proceeds much faster for the 7:3-bin system than for the remaining ones, the most striking difference is in the final mass fraction of N. As predicted in Table~\ref{tab:case_2_reservoir_conditions}, the 7:3-bin system reaches equilibrium with $y_\mathrm{N}^\mathrm{eq} = 0.335$, while the reference value for the full set of levels is $y_\mathrm{N}^\mathrm{eq} = 0.271$. Once the number of bins is increased by a factor of ten, to 70:30 (dark blue dashed curves in Fig.~\ref{fig:master_equation_results_old_bins_lower_temp}) or more, the mismatch between the binned system and the reference profiles becomes much less noticeable. By comparison, the alternative binning shown in Fig.~\ref{fig:master_equation_results_new_bins_lower_temp}, exhibits much closer agreement with the reference curves for all combinations tested. Even the smallest 9:1-system follows the other curves fairly closely, and its final equilibrium state only differs by a few percent from the reference values. This suggests that at the low-temperature conditions the alternative binning provides the best approximation to the full RVC model with the smallest number of bins.


Results for the medium-temperature case are shown in Figs.~\ref{fig:master_equation_results_old_bins} (original binning) and \ref{fig:master_equation_results_new_bins} (alternative binning) respectively. Compared to the low-temperature, low-pressure initial conditions of Figs.~\ref{fig:master_equation_results_old_bins_lower_temp} and \ref{fig:master_equation_results_new_bins_lower_temp}, reaching the final equilibrium state now only takes about 1/1000~\textsuperscript{th} of the time, due to an initial reservoir pressure approximately 5 times greater. Apart from the difference in relaxation times, the mass fraction- and temperature profiles exhibit slight differences with respect to the low-temperature cases. First, relative to the low-temperature case dissociation now is triggered at an earlier stage and overlaps more with the initial relaxation of translational and rovibrational modes. This can be inferred from the fact that the internal temperature reaches its maximum value and begins to decrease before it equalizes with the translational mode temperature. This is in contrast with the lower-temperature case, where the maximum internal temperature is reached at exactly the point where it meets the translational temperature curve, and before a significant rise in atomic nitrogen concentration can be observed. Of course, given this higher initial translational temperature, there is more energy per unit mass stored in the system, causing the higher equilibrium temperature of 5693~K, but also shifting the equilibrium gas composition towards a greater mass fraction of atomic nitrogen.

As before, we can see that the equally-spaced, 7:3-bin case is the worst at reproducing the behavior of the full RVC model, exhibiting much faster dissociation, but also failing to come close to the correct final equilibrium composition (for the 7:3-bins $y_\mathrm{N}^\mathrm{eq} = 0.782$, while the reference value for the full set of levels is $y_\mathrm{N}^\mathrm{eq} = 0.702$). Again, this discrepancy becomes less severe as the number of bins is increased to 70:30 and higher, and again, the alternative binning does a slightly better job of following the behavior the full system.

More differences between the low-temperature and medium-temperature can be observed. Whereas in Figs.~\ref{fig:master_equation_results_old_bins_lower_temp} and \ref{fig:master_equation_results_new_bins_lower_temp} the binned systems would both relax and dissociate at a slightly faster pace than the full RCV model, the same is not true for the medium-temperature conditions. In Figs.~\ref{fig:master_equation_results_old_bins} and \ref{fig:master_equation_results_new_bins} we can see the temperature profiles of the binned systems initially lagging behind the reference curves. They then overtake the reference curves at $t \approx 10^{-5} \, \mathrm{sec}$, near the point where $T_\mathrm{int}$ reaches its maximum. Simultaneously, dissociation is seen to proceed at a slightly slower rate for the binned systems, but then overtakes the reference curves at approximately the same time.


Finally, Figs.~\ref{fig:master_equation_results_old_bins_higher_temp} and \ref{fig:master_equation_results_new_bins_higher_temp} show the results for the high-temperature case. Given the much higher initial pressure and associated collision rate, the gas in the reactor now only takes about $10^{-4}$ seconds to reach its equilibrium. Compared to the previous two cases, there are several noticeable differences. First, due to the much higher initial translational temperature, the gas ends up almost fully dissociated, with an equilibrium temperature slightly above 20000 kelvin. In fact, when given to a precision of three significant digits, Table~\ref{tab:case_2_reservoir_conditions} reports the final equilibrium to be $y_\mathrm{N}^\mathrm{eq} = 1.00$ regardless of the model used. 

Second, relaxation of the translational and rovibrational modes towards a common reservoir temperature now takes place concurrently with $\mathrm{N_2}$-dissociation, and in relative terms is completed at a much later stage than in the previous cases. This is clearly visible in the temperature profiles of Figs.~\ref{fig:master_equation_results_old_bins_higher_temp}(b) and \ref{fig:master_equation_results_new_bins_higher_temp}(b), where it takes about $7 \times 10^{-5}$ seconds for the translational- and rovibrational mode temperatures in the full system (solid black curves) to reach a common value. At this stage, the gas is already almost fully dissociated, and the few remaining $\mathrm{N_2}$-molecules require many more collisions to reach equilibrium populations.

However, the most striking difference with respect to the lower-temperature cases is in the behavior of the binned systems with respect to the full RVC model. Focusing on Fig.~\ref{fig:master_equation_results_old_bins_higher_temp}(a) first, we can see that with the original, equally-spaced bins the mass fraction profiles for all three binned systems clearly lag behind those of the full system, up until $t \approx 2 \times 10^{-6} \, \mathrm{sec}$. It is only after this time that these trends are reversed. As the number of bins is increased, the differences become less noticeable, but they do not disappear entirely. At the same time, Fig.~\ref{fig:master_equation_results_old_bins_higher_temp}(b) reveals that the translational- and internal temperature profiles of the binned systems also lag behind the full model by a significant margin. As could be expected, the worst prediction is given by the 7:3-bin system. In addition to being the least accurate in terms of time-history, it also over-predicts the final equilibrium temperature by about 16~\%. Using 70:30 bins improves prediction of the equilibrium state, but does not seem to have a major effect on the time history of the mass fraction- or temperature profiles. Even with the 700:269-bin system, we are still off by a noticeable margin. 

Unfortunately, this situation is only slightly improved by using the alternative, variably-spaced bins. As can be seen in Fig.~\ref{fig:master_equation_results_new_bins_higher_temp}, the initial lag of the binned systems behind the reference profiles is of similar magnitude as it was with the original bins. This is true for both the species mass fractions and the temperature profiles. Only at the later stages, i.e. for $t > 2 \times 10^{-6} \, \mathrm{sec}$, does the agreement between the binned system profiles and the reference RVC curves begin to improve. The greatest advantage of the variably-sized bins is that the thermodynamic properties are better reproduced with as few as 10 bins, which is consistent with the results from the lower-temperature cases.


\section{Integration of bin-specific reaction cross sections into DSMC collision algorithm} \label{sec:bin_model_dsmc_integration}

After studying the two alternative binning strategies using master equation simulations in Sec.~\ref{sec:parametric_study_kinetics}, we now move on to describing the integration of the URVC bin model into the DSMC framework. We will first describe the manner in which the state-resolved cross sections for internal energy exchange and dissociation are integrated into the DSMC collision routines, so as to reproduce the correct macroscopic reaction rates. We also discuss how the post-collision velocities of the departing particles are determined, enforcing conservation of momentum and total collision energy along the way. Finally, we verify our DSMC implementation by comparing our results to the master equation calculations discussed in Sec.~\ref{sec:parametric_study_kinetics}.

Several modifications have to be made to a DSMC code in order to integrate the detailed chemistry model described in our previous publication~\cite{torres18a}. In that paper we discussed our method for extracting state-resolved coarse-grain cross sections from the Ames database. Here we will concentrate on the implementation issues that arise while integrating the rovibrational bin model into DSMC. The most important additions to the DSMC algorithm concern the collision routines. Only the modifications necessary to integrate the URVC bin model for the N3-system are mentioned here. For simplicity this discussion will assume that Bird's No-Time-Counter (NTC) scheme, as discussed in~\cite{bird94a}, is being used. 

Before outlining the collision algorithm itself, two relevant points will be discussed. The first one concerns the determination of the state-specific total cross section in Sec.~\ref{sec:state_specific_total_cross_section}. The second point is related and concerns the determination of the state-specific reaction probabilities, which are introduced in Sec.~\ref{sec:state_specific_reaction_probabilities}.

\subsection{State-specific total cross section for \texorpdfstring{$\boldsymbol{\mathrm{N_2} \left( k \right)}$+$\boldsymbol{\mathrm{N}}$}{N2(k)+N}} \label{sec:state_specific_total_cross_section}

The total integrated collision cross section $\bar{\sigma}_k^T$ of the collision pair $\mathrm{N_2}\left( k \right) + \mathrm{N}$ is calculated as the sum over all inelastic, intra-bin and reactive cross sections for this particular pseudo-species pairing:
\begin{equation}
 \bar{\sigma}_{k}^T = \sum_{l \in \mathcal{I}_k} \bar{\sigma}_{k \rightarrow l}^E + \bar{\sigma}_k^{Df}, \qquad k \in \mathcal{K}_\mathrm{BP}.
 \label{eq:bin_specific_total_cross_section}
\end{equation}

Here, $\bar{\sigma}_{k \rightarrow l}^E$ represents the full set of bin-specific excitation, deexcitation and intra-bin (i.e. $\bar{\sigma}_{k \rightarrow k}^\mathrm{coll} = \bar{\sigma}_{k \rightarrow (l = k)}^\mathrm{E}$) cross sections from pre-collision bin $\mathrm{N_2}\left(k\right)$ to all possible post-collision bins $\mathrm{N_2}\left(l\right)$. In addition, $\bar{\sigma}_k^{Df}$ is the cross section for dissociation from bin $\mathrm{N_2}\left(k\right)$ when colliding with $\mathrm{N}$.

\begin{figure}
 \centering
 \includegraphics[width=\columnwidth]{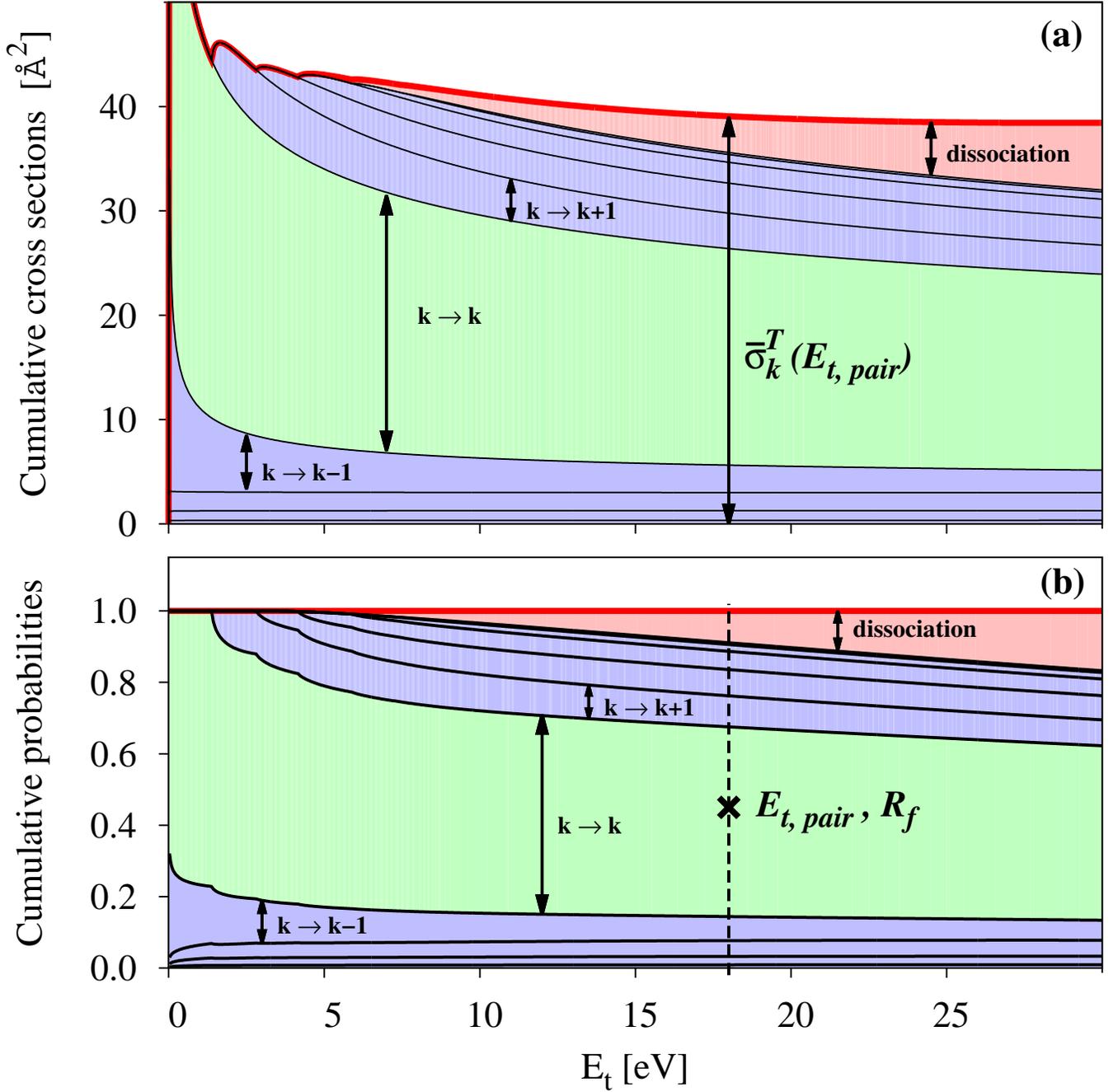}
 
 \caption{Cumulative state-specific cross sections and probabilities for collision pair $\mathrm{N_2} \left( k \right) + \mathrm{N}$, as a function of relative translational energy $E_t = \frac{1}{2} \, \mu_\mathrm{N_2,N} \, g^2$. The state of any particular collision pair can be represented graphically as a specific point at the coordinates ($E_{t\mathrm{pair}}$, $R_f$). The outcome of the collision undergone by the pair corresponds to the region in the probability graph on which this point lies.} \label{fig:cumulative_cross_sections_and_probabilities_original}
\end{figure}

The sum expressed in Eq.~(\ref{eq:bin_specific_total_cross_section}) is represented graphically in Fig.~\ref{fig:cumulative_cross_sections_and_probabilities_original}(a). Each possible outcome of the collision has its own integrated cross section, which is represented by one of the shaded areas. For example, the deexcitation reaction to the adjacent lower bin, i.e. $\mathrm{N_2} \left( k \right) + \mathrm{N} \rightarrow \mathrm{N_2} \left( k - 1 \right) + \mathrm{N}$ is labeled as $k \rightarrow k - 1$ and is shaded in blue. At a given relative translational energy $E_t$, the cross section for this process is then given by the ``height'' of the corresponding shaded area (indicated by the double-headed arrows). As can be inferred from Fig.~\ref{fig:cumulative_cross_sections_and_probabilities_original}(a), the individual cross sections for all possible outcomes have been ``stacked on top of each other'' in sequence of the post-collision bin $l$. Thus, immediately above $k \rightarrow k - 1$ follows the special case where the post-collision bin remains unchanged, i.e. $k \rightarrow k$, which is shaded in green. Further cycling through all possible post-collision outcomes, the following cross sections all represent excitation processes to higher-energy bins  $l > k$ and are again shaded in blue. The final cross section is the one for dissociation, i.e. $\mathrm{N_2} \left( k \right) + \mathrm{N} \rightarrow 3 \, \mathrm{N}$, which is shaded in red. The overall sum of all these contributions yields the total collision cross section for the pair $\mathrm{N_2} \left( k \right) + \mathrm{N}$, which is represented in Fig.~\ref{fig:cumulative_cross_sections_and_probabilities_original}(a) by the thick red line.

Notice in the definition of Eq.~(\ref{eq:bin_specific_total_cross_section}) the dependence on parameter $k$, which implies the fact that a different total cross section exists for each pre-collision pseudo-species pairing $\mathrm{N_2} \left( k \right) + \mathrm{N}$. This is in contrast with most conventional DSMC implementations, where it is usually assumed that the total cross section for a given species pair is independent of the internal energy states of the colliding particles (e.g. when using the VHS/VSS models). The dependence of the total cross section on the pre-collision internal energy states was also recognized in previous state-to-state DSMC implementations of the NASA N3 database~\cite{kim14a}.

The total cross section is needed to determine the acceptance probability of a given collision pair in Bird's NTC scheme (see Sec.~\ref{sec:ntc_acceptance_probability}), but is also necessary for determination of the state-specific reaction probabilities discussed in Sec.~\ref{sec:state_specific_reaction_probabilities}.

\subsection{State-specific reaction probabilities for \texorpdfstring{$\boldsymbol{\mathrm{N_2} \left( k \right)}$+$\boldsymbol{\mathrm{N}}$}{N2(k)+N}} \label{sec:state_specific_reaction_probabilities}

The relative frequencies (or probabilities) of the collision yielding a particular post-collision outcome are given by the ratio of each individual cross section to the pair's total cross section. Thus for excitation/deexcitation reactions $P_{k \rightarrow l}^E = \bar{\sigma}_{k \rightarrow l}^E / \bar{\sigma}_k^T$ for all $k, l \in \mathcal{K}_\mathrm{BP}$ and for dissociation $P_{k}^{Df} = \bar{\sigma}_{k}^{Df} / \bar{\sigma}_k^T$ for all $k \in \mathcal{K}_\mathrm{BP}$.

Adding up the individual terms for all post-collision outcomes one-by-one in the same sequence as the cross sections in Eq.~(\ref{eq:bin_specific_total_cross_section}) yields the corresponding cumulative probabilities. They can be represented in an array $P_k^\mathrm{\Sigma}$, which is constructed in the following manner:
\begin{equation}
 P_k^\mathrm{\Sigma} = \left[
 \begin{array}{l}
  P_{k \rightarrow 1}^E \\
  P_{k \rightarrow 1}^E + P_{k \rightarrow 2}^E \\
  \qquad \vdots \\
  P_{k \rightarrow 1}^E + P_{k \rightarrow 2}^E + \, \cdots \,  + P_{k \rightarrow n_\mathrm{bins}}^E \\
  P_{k \rightarrow 1}^E + P_{k \rightarrow 2}^E + \, \cdots \,  + P_{k \rightarrow n_\mathrm{bins}}^E + P_{k}^{Df}
 \end{array} \right]. \label{eq:cumulative_probabilities_bin}
\end{equation}

This array has length $n = n_\mathrm{bins} + 1$, each row corresponding to one of the possible outcomes available to the collision pair $\mathrm{N_2} \left( k \right)$+$\mathrm{N}$. Notice that for the last element $P_k^\mathrm{\Sigma} \overset{!}{=} 1$, since it is calculated as the sum over all individual probabilities. As an example Fig.~\ref{fig:cumulative_cross_sections_and_probabilities_original}(b) shows the cumulative probabilities corresponding to the state-specific cross sections of Fig.~\ref{fig:cumulative_cross_sections_and_probabilities_original}(a). Again, the individual probabilities of all transitions are stacked on top of each other, with the ``height'' of each shaded strip representing the probability of that particular outcome. The cumulative probabilities of Eq.~(\ref{eq:cumulative_probabilities_bin}) are represented by the solid dividing lines between the shaded areas. Although not strictly necessary, the definition given by Eq.~(\ref{eq:cumulative_probabilities_bin}) is introduced here, because as will be seen in Sec.~\ref{sec:ntc_post_collision_outcome}, it is a convenient device for determining the post-collision outcome in the DSMC routines.

\subsection{Collision acceptance probability of \texorpdfstring{$\boldsymbol{\mathrm{N_2} \left( k \right)}$+$\boldsymbol{\mathrm{N}}$}{N2(k)+N}} \label{sec:ntc_acceptance_probability}

Recall that according to the NTC method, in a particular collision cell with a total number of particles $N_P$, a particle weighting factor $W_P$, a time step size $\Delta t$ and cell volume $V_\mathrm{cell}$, the number of pairs to be \emph{tested} for collision is given by:
\begin{equation}
 N_\mathrm{pairs} = \frac{N_P \left(N_P - 1\right)}{2} \left[ \sigma^T \left( g \right)  \cdot g \right]_\mathrm{max} \frac{W_P \, \Delta t}{V_\mathrm{cell}}, \label{eq:ntc_method_number_of_pairs}
\end{equation}
where $[ \sigma^T \left( g \right)  \cdot g ]_\mathrm{max}$ is the cell-specific maximum of $[ \sigma^T \left( g \right) \cdot g ]_\mathrm{pair}$ recorded over all previously tested collision pairs, regardless of the species of the particular collision partners. 
For all pairs which are composed of one $\mathrm{N_2}\left(k\right)$-molecule and one $\mathrm{N}$-atom at the particular relative translational energy $E_\mathrm{pair} = \frac{1}{2} \, \mu_{\mathrm{N_2}, \mathrm{N}} \, g_\mathrm{pair}^2$, the \emph{probability of acceptance} is given by the ratio:
\begin{equation}
 P_\mathrm{pair} = \frac{\left[ \sigma_k^T \left( g \right) \cdot g \right]_\mathrm{pair} }{ \left[ \sigma^T \left( g \right) \cdot g \right]_\mathrm{max} }. \label{eq:ntc_acceptance_probability}
\end{equation}

For every such random pair the total cross section must then be evaluated according to Eq.~(\ref{eq:bin_specific_total_cross_section}) at the pair's specific relative energy $E_\mathrm{pair}$. As usual in the NTC method, the pair is accepted for collision only if $P_\mathrm{pair} > R_f$, where $R_f$ is a random real number in the interval $\left[0,1\right]$. 

During the remainder of the pair acceptance/rejection phase this process has to be repeated independently for every other random pair, which happens to involve $\mathrm{N_2}\left(k\right)$ and $\mathrm{N}$ as collision partners. By affecting the acceptance probabilities of individual pairs, the total cross section of Eq.~(\ref{eq:bin_specific_total_cross_section}) directly influences the collision rate between pseudo-species $\mathrm{N_2}\left(k\right)$ and $\mathrm{N}$. 

\subsection{Determination of post-collision state of \texorpdfstring{$\boldsymbol{\mathrm{N_2} \left( k \right)}$+$\boldsymbol{\mathrm{N}}$}{N2(k)+N}}
\label{sec:ntc_post_collision_outcome}

\begin{figure}
 \centering
 \includegraphics[width=0.99\columnwidth]{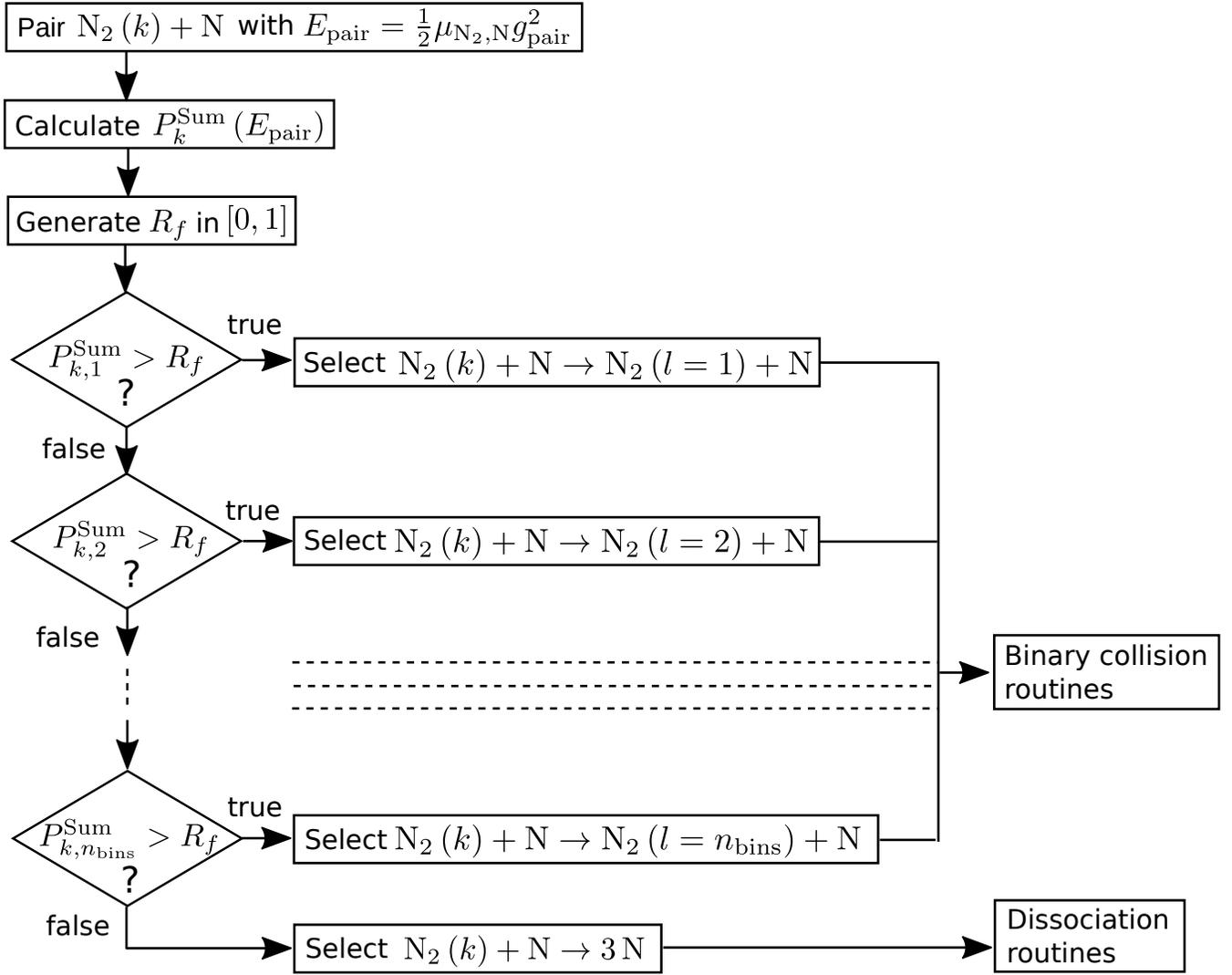}

 \caption{Algorithm for randomly selecting a post-collision internal state (i.e. bin index $l$, or dissociation event) for a given $\mathrm{N_2} \left( k \right)$+$\mathrm{N}$-pair after having been accepted for collision with the NTC scheme. This procedure is repeated for every accepted $\mathrm{N_2} \left( k \right)$+$\mathrm{N}$-pair in the DSMC cell}
 \label{fig:bin_outcome_selection}
\end{figure}

Once all $N_\mathrm{pairs}$ potential collision pairs have been flagged either for acceptance or rejection, a second loop is performed over the accepted pairs. If a given $\mathrm{N_2} \left( k \right)$+$\mathrm{N}$-pair has been accepted for collision, the following task is to randomly select its post-collision outcome in accordance with its state-specific reaction probabilities. This is where the set of cumulative probabilities as defined in Eq.~(\ref{eq:cumulative_probabilities_bin}) becomes useful, since selecting the post-collision outcome is straightforward by generating a random fraction $R_f$ in the interval $\left[ 0, 1 \right]$ and comparing it in sequence to each element in vector $P_k^\mathrm{\Sigma}$. The collision outcome whose cumulative probability is closest to, but still greater than $R_f$ will be chosen. This algorithm can be represented by the flow chart shown in Fig.~\ref{fig:bin_outcome_selection}. Notice that if the random number turns out to be greater than all but the last element of $P_k^\mathrm{\Sigma}$, the dissociation reaction is chosen by default, since the last cumulative probability in the array is equal to one by design. Alternatively, a graphical interpretation of the procedure is to plot the point $\left( E_\mathrm{pair}, R_f \right)$ on the graph of Fig.~\ref{fig:cumulative_cross_sections_and_probabilities_original}(b) and to select the post-collision outcome according to the particular interval in which the point lies.

Once a given collision outcome has been selected, the remainder of the procedure depends on whether the $\mathrm{N_2}$-molecule remains intact, or whether dissociation occurs. These two different paths, designated in Fig.~\ref{fig:bin_outcome_selection} as ``binary collision routines'' and ``dissociation routines'' will now be explained.

\subsubsection{Binary collision routines} \label{sec:binary_collision_routines}

If the $\mathrm{N_2} \left( k \right)$+$\mathrm{N}$-pair transitions to a new post-collision bin, it is only necessary to update the molecule's pseudo-species label, reflecting the fact that it has transitioned to state $\mathrm{N_2}\left(l\right)$, as well as to determine the post-collision relative speed $g_{12}^\prime = \left| \boldsymbol{c}_1^\prime - \boldsymbol{c}_2^\prime \right|$ of the pair (for the labeling, refer to Fig.~\ref{fig:n2n_inelastic_processes_sketches}\subref{fig:n2n_inelastic_collision_relative_velocities_bins}:
\begin{equation}
 g_{12}^\prime = \sqrt{g_{12}^2 + 2 \left( \bar{E}_k - \bar{E}_l \right) / \mu_\mathrm{N_2,N}}.
\end{equation}
Once this value is known, the post-collision velocities of the $\mathrm{N_2} \left( l \right)$+$\mathrm{N}$-pair are determined assuming isotropic scattering at $g_{12}^\prime$, exactly as would be the case for an elastic hard-sphere collision. The main reason for assuming isotropic scattering is that the NASA N3 database contains only cross section data integrated over all deflection angles. This means that all of the information on the possible anisotropy of the binary collisions has been lost. This is a consequence of these data having been originally compiled with hydrodynamic methods in mind, where this information is of secondary importance.

\subsubsection{Dissociation routines} \label{sec:dissociation_routines}

\begin{figure}
 \centering
 \subfloat[Binary inelastic collision between $\mathrm{N_2} \left( k \right)$ and $\mathrm{N}$]{
  \includegraphics[width=0.52\columnwidth]{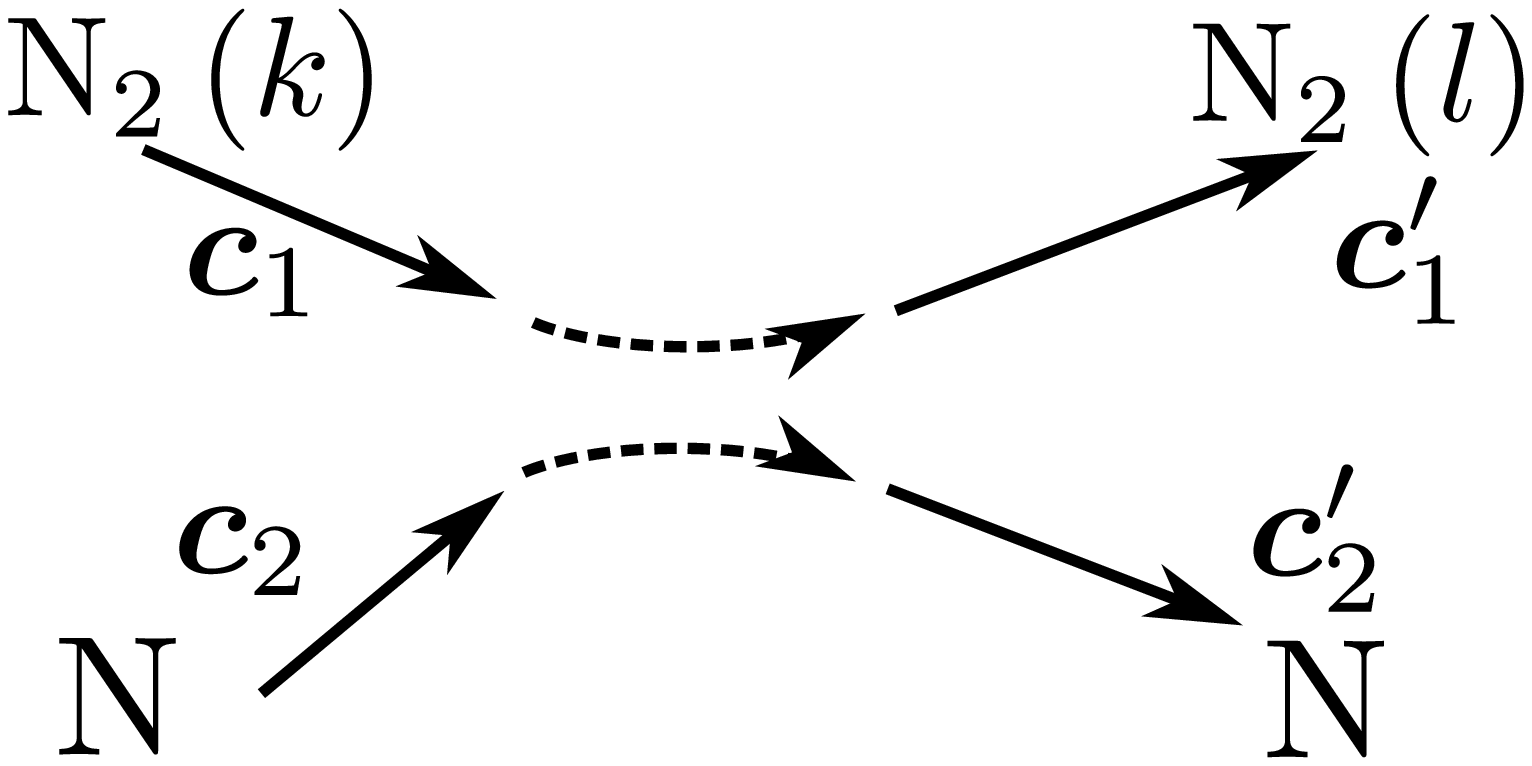}
  \label{fig:n2n_inelastic_collision_relative_velocities_bins}}

 \subfloat[Dissociation of $\mathrm{N_2} \left(k\right)$ by collision with $\mathrm{N}$]{
  \includegraphics[width=0.64\columnwidth]{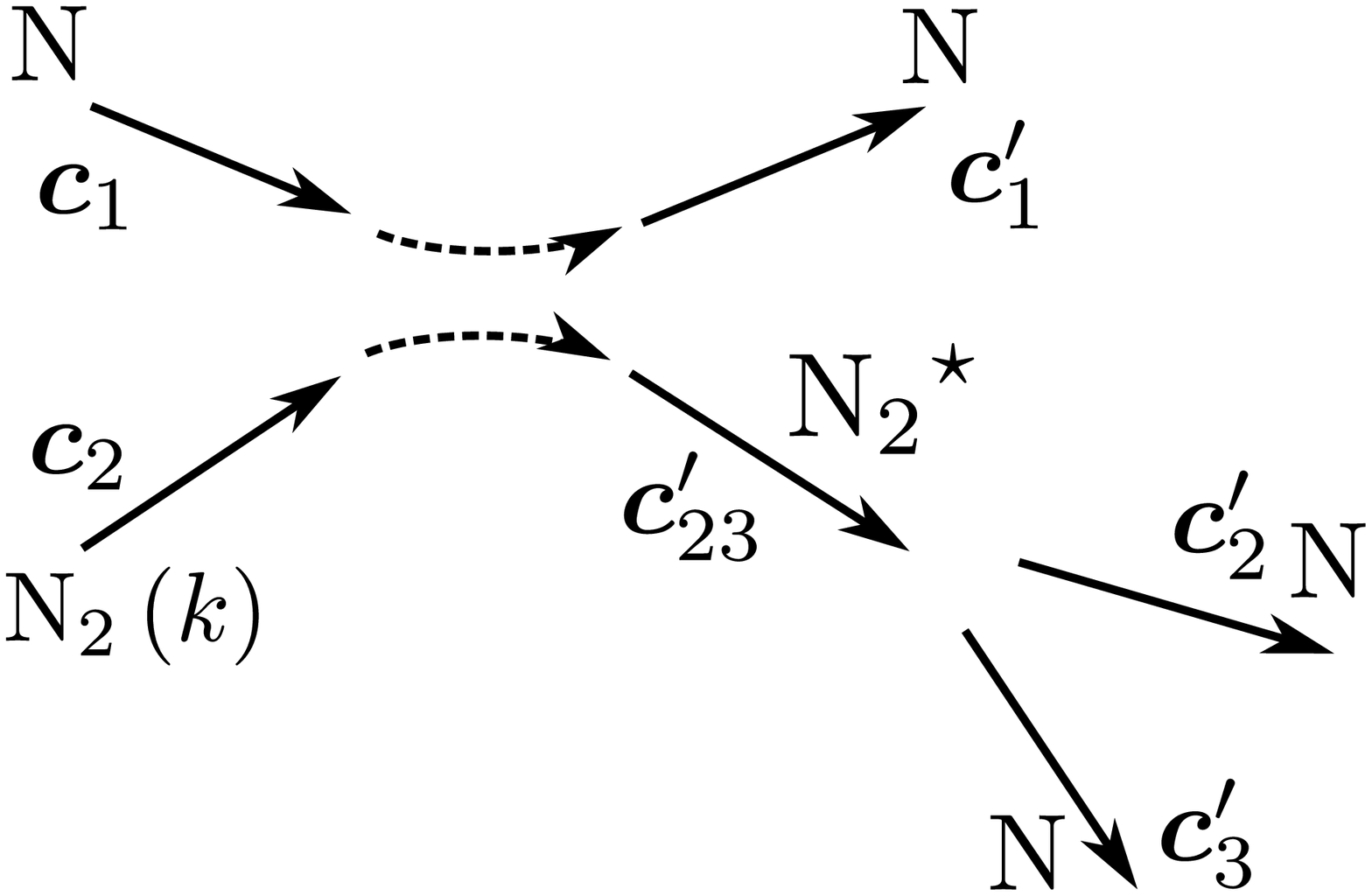}
  \label{fig:n2n_dissociation_reaction_relative_velocities_bins}}

 \caption{Sketch of collision processes and definition of pre- and post-collision velocities}
 \label{fig:n2n_inelastic_processes_sketches}
\end{figure}

If dissociation is selected, the remaining procedure is slightly more involved. In order to better visualize the sequence of events and the labeling of each particle, refer to Fig.~\ref{fig:n2n_inelastic_processes_sketches}\subref{fig:n2n_dissociation_reaction_relative_velocities_bins}. The procedure is split into two parts. In the first phase the $\mathrm{N_2}$-particle, which will later split up into two N-atoms, is labeled inside the code as an ``activated complex'' $\mathrm{N_2^\star}$, and a regular binary collision between this intermediate particle and the original nitrogen atom is performed. Then, after all collision pairs in the cell have been processed in the regular manner, a loop over all $\mathrm{N_2}$-particles which had previously been labeled as activated complexes, is performed in order to split them into atoms. The whole procedure is performed in such a manner as to conserve momentum and overall energy in each collision. The first phase proceeds as follows:
\begin{enumerate}
 \item Sample the post-collision relative translational energy of the $\mathrm{N_2^\star}$+$\mathrm{N}$-pair from a uniform distribution in the interval $\left[0,E_\mathrm{max}\right]$, i.e. $E_{1\left(23\right)}^\prime = E_\mathrm{max} \cdot R_f$, where again $R_f$ is a random number sampled from $\left[0, 1\right]$.
 The upper limit $E_\mathrm{max}$ constitutes the maximum available energy to be distributed among the collision products after the energy of formation of the two new nitrogen atoms, i.e. $2 E_\mathrm{N}$, has been deducted:
 \begin{equation}
  E_\mathrm{max} = \frac{1}{2} \, \mu_\mathrm{N_2,N} \, g_{12}^2 - \left( 2 E_\mathrm{N} - \bar{E}_k \right).
 \end{equation}
 The remaining collision energy, which is associated with the activated complex, is calculated as $E^\star = E_\mathrm{max} - E_{1\left(23\right)}^\prime$ and is kept in memory for use in the second phase.
 \item Using $g_{1\left(23\right)}^\prime = \sqrt{2 \, E_{1\left(23\right)}^\prime / \mu_\mathrm{N_2,N}}$, determine velocity $\boldsymbol{c}_1^\prime$ of the original nitrogen atom and $\boldsymbol{c}_{23}^\prime$ of the activated complex, assuming isotropic scattering between the two.
\end{enumerate}
Thus, in the first phase all $\mathrm{N_2}$-molecules labeled for dissociation will first undergo regular binary collisions, which yield the particles $\mathrm{N_2^\star}$ and $\mathrm{N}$ with respective velocities $\boldsymbol{c}_{23}^\prime$ and $\boldsymbol{c}_1^\prime$. Apart from removing $E^\star$ from the pre-collision collision energy and the re-labeling of the dissociating molecules as $\mathrm{N_2^\star}$, this phase is identical to the procedure followed for collisions between inert species.
The second phase involves only the activated-complex-particles:
\begin{enumerate}
 \item For each given $\mathrm{N_2^\star}$, re-assign its velocity $\boldsymbol{c}_{23}^\prime$ to be the center-of-mass velocity of the two $\mathrm{N}$-atoms which will be created.
 \item Retrieve $E^\star$ from memory and calculate $g_{23}^\prime = \sqrt{2 E^\star / \mu_{\mathrm{N,N}}}$, which will become the post-collision relative speed between the two atoms.
 \item Re-label the activated complex particle as an $\mathrm{N}$-atom and generate a new particle in memory to be the third $\mathrm{N}$-atom.
 \item Based on $g_{23}^\prime$ determine the velocities $\boldsymbol{c}_2^\prime$ and $\boldsymbol{c}_3^\prime$ of both $\mathrm{N}$-atoms assuming isotropic scattering.
\end{enumerate}
It should be noted that, although this procedure conserves momentum and energy in each collision, the post-collision velocities $\boldsymbol{c}_1^\prime$,  $\boldsymbol{c}_2^\prime$ and  $\boldsymbol{c}_3^\prime$ determined via this algorithm are somewhat arbitrary. They come about as a direct result of the assumed uniform distribution of the post-collision energy $E_\mathrm{max}$ between the between the three particles and by assuming isotropic scattering in each case. Again, this is a choice which had to be made due to the lack of more precise information on the deflection angles in the original N3 database.

Before moving on to Sec.~\ref{sec:verification}, it should be mentioned that recombination reactions of the type $\mathrm{N}+\mathrm{N}+\mathrm{N} \rightarrow \mathrm{N_2}\left(k\right) +\mathrm{N}$ were not implemented in the current state-to-state chemistry algorithm. Although it may seem trivial to ``invert'' the arrows in Fig.~\ref{fig:n2n_inelastic_processes_sketches}\subref{fig:n2n_dissociation_reaction_relative_velocities_bins}, so as to reverse the sense of the reaction, devising the correct DSMC algorithm for this process is not straightforward. The first problem is the one of determining the correct three-particle collision rate, based on the particles in a given DSMC cell. No rigorous approach for this problem seems to have been proposed so far. The second difficulty lies in the lack of knowledge of the state-specific recombination cross sections, which must be consistent with the corresponding dissociation cross sections to ensure detailed balance. In fact, at the moment it is not clear if an unambiguous definition of such cross sections is even possible. Since the number of degrees of freedom of the three-particle system is greater than that of the resulting particle pair after recombination, the choice of parameters by means of which one defines a particular collision does not seem to be unique~\cite{nagnibeda09a} and a consistent recombination algorithm can not be presented at the moment.


\section{Verification of DSMC implementation} \label{sec:verification}

We now present the verification of our DSMC model in a code written for this express purpose. We study the model's behavior with the help of the same test case discussed in Sec.~\ref{sec:parametric_study_kinetics} and compare the time evolution of the DSMC-derived flow macroparameters to the ones obtained from the equivalent master-equation solution. Before discussing the results of the simulations, we have to mention two major compromises in our physical modeling. In Sec.~\ref{sec:bin_model_dsmc_integration} we discussed at some length the manner in which the reaction mechanism for $\mathrm{N_2}\left(k\right)$+$\mathrm{N}$-collision pairs was integrated into the DSMC collision routines. For these species pairings we only needed the bin-specific cross sections, previously derived from the Ames N3 database~\cite{torres18a}. However, at no time did we go into detail about the other species pairings that must be naturally occurring in an $\mathrm{N_2}$,$\mathrm{N}$-mixture. 

The most prevalent one in weakly dissociated nitrogen would be between two molecules, possibly populating distinct pre-collision rovibrational states. In the context of our bin model, such collisions can result in excitation/deexcitation reactions with either one, or both molecules' internal energies changing, or alternatively in collisions with no internal energy change. Furthermore, single- or possibly double dissociation in molecule-molecule encounters is the mechanism providing the initial seed atomic nitrogen, which is necessary to enable any of the N3-reactions in the first place. Admittedly, the set of possible reactions for this N4-system is much larger than the one for N3, and constitutes a significant part of a complete state-to-state model for nitrogen. However, a lack of appropriate cross section data prevented us from including them it in our current model. Out of necessity, whenever two $\mathrm{N_2}$-molecules are selected for collision in our DSMC implementation, we assume that an elastic, hard-sphere collision takes place. Once even a small amount atomic nitrogen is available in the gas, N3-encounters are several orders of magnitude more effective than N4-encounters in promoting inelastic collision processes.

The second species pairing, more prevalent in strongly dissociated nitrogen, involves two N-atoms. For the purpose of our current model, we have assumed that atomic nitrogen only exists in its ground electronic state. Therefore, within the limits of our current model atom-atom encounters can only result in elastic scattering. In Table~\ref{tab:bin_model_vhs_parameters} we list the VHS parameters used in determining the elastic cross sections, adapted from Stephani et al~\cite{stephani12a}. We are aware that the listed VHS parameters are tuned to a much lower reference temperature than the conditions of our test case. Since we also assume isotropic, i.e. hard-sphere, scattering for all binary encounters, we accept that we will not be able to accurately reproduce macroscopic transport properties, such as diffusion coefficients and thermal conductivities in our multi-component mixture. However, since in the current paper we are primarily concerned with studying the state-to-state kinetics, the precise transport properties of the gas are of secondary importance. In any case, in the space-homogeneous heat bath calculations shown in this paper transport phenomena play no role.

\begin{table}
 \centering
 \caption{VHS parameters used in elastic collisions} \label{tab:bin_model_vhs_parameters}
 
 \begin{tabular}{c c c c}
                                & $d_\mathrm{ref}$ [\AA] & $\omega$ & $T_\mathrm{ref}$ $[\mathrm{K}]$ \\ \hline
  $\mathrm{N_2}$-$\mathrm{N_2}$ & $3.20$ & $0.680$ & $2880$ \\
  $\mathrm{N}$-$\mathrm{N}$ & $2.60$ & $0.700$ & $2880$
 \end{tabular}
\end{table}

The second major simplification in our implementation was already mentioned at the end of Sec.~\ref{sec:ntc_post_collision_outcome}, and concerns neglecting recombination reactions. We discuss the implications of neglecting recombination in the DSMC model in Sec.~\ref{sec:neglecting_recombination}.

\subsection{DSMC verification test case: Adiabatic reactor} \label{sec:dsmc_verification_test_case}

We simulated the test case of Sec.~\ref{sec:parametric_study_kinetics} with our DSMC code at the medium-temperature initial conditions (second row of Table~\ref{tab:case_2_reservoir_conditions}). Since this test case involves no spatial variations in flow macroparameters, a single DSMC cell is enough to represent this adiabatic, isochoric reactor. We generated initial velocity components for all DSMC particles following Maxwellian distributions at $T_\mathrm{t}^0 = 62\,546 \, \mathrm{K}$ and a static pressure of $3164 \, \mathrm{Pa}$, while the initial bin populations of $\mathrm{N_2}$ follow a Boltzmann distribution at $T_\mathrm{int}^0 = 300$ K. A small fraction of the initial particles are generated as atomic nitrogen (mass fraction: $y_\mathrm{N}^0 = 0.014$), whereas the rest are $\mathrm{N_2}$-molecules distributed among the different bins. Just as was the case in the master equation calculations, this atomic nitrogen is necessary to trigger the processes of internal energy exchange and dissociation.

In each of the cases studied, the gas mixture is initially represented by approximately $95\,000$ DSMC simulator particles. As the simulation progresses, this number gradually increases to about $160\,000$, due to the dissociation of $\mathrm{N_2}$. A constant time step $\Delta t = 10^{-8} \, \mathrm{s}$ was chosen, which is well below the mean collision time for these conditions. All simulations were run on the same workstation and the instantaneous macroscopic samples are ensemble-averages over four independent simulations. Thus, the macroscopic samples derived from our DSMC simulations are all based on a total of approximately $380\,000 - 640\,000$ particles. Recall that Table~\ref{tab:case_2_reservoir_conditions} summarizes both the initial and final states of the reservoir. Although not all cases are listed, this same test case was simulated with DSMC using $n_\mathrm{bins} = 10, 20, 100$ and $200$. As in the master equation study of Sec.~\ref{sec:parametric_study_kinetics}, we used both the original, equally-spaced bins with a bound-to-predissociated bin ratio of B:P=7:3, as well as the alternative, variably-spaced bins with B:P=9:1. In all our simulated cases very close agreement between the DSMC macroparameters and the corresponding master equation results was observed. Therefore, here we show only a few representative cases for verification purposes.

\begin{figure}
 \centering 
 \includegraphics[width=\columnwidth]{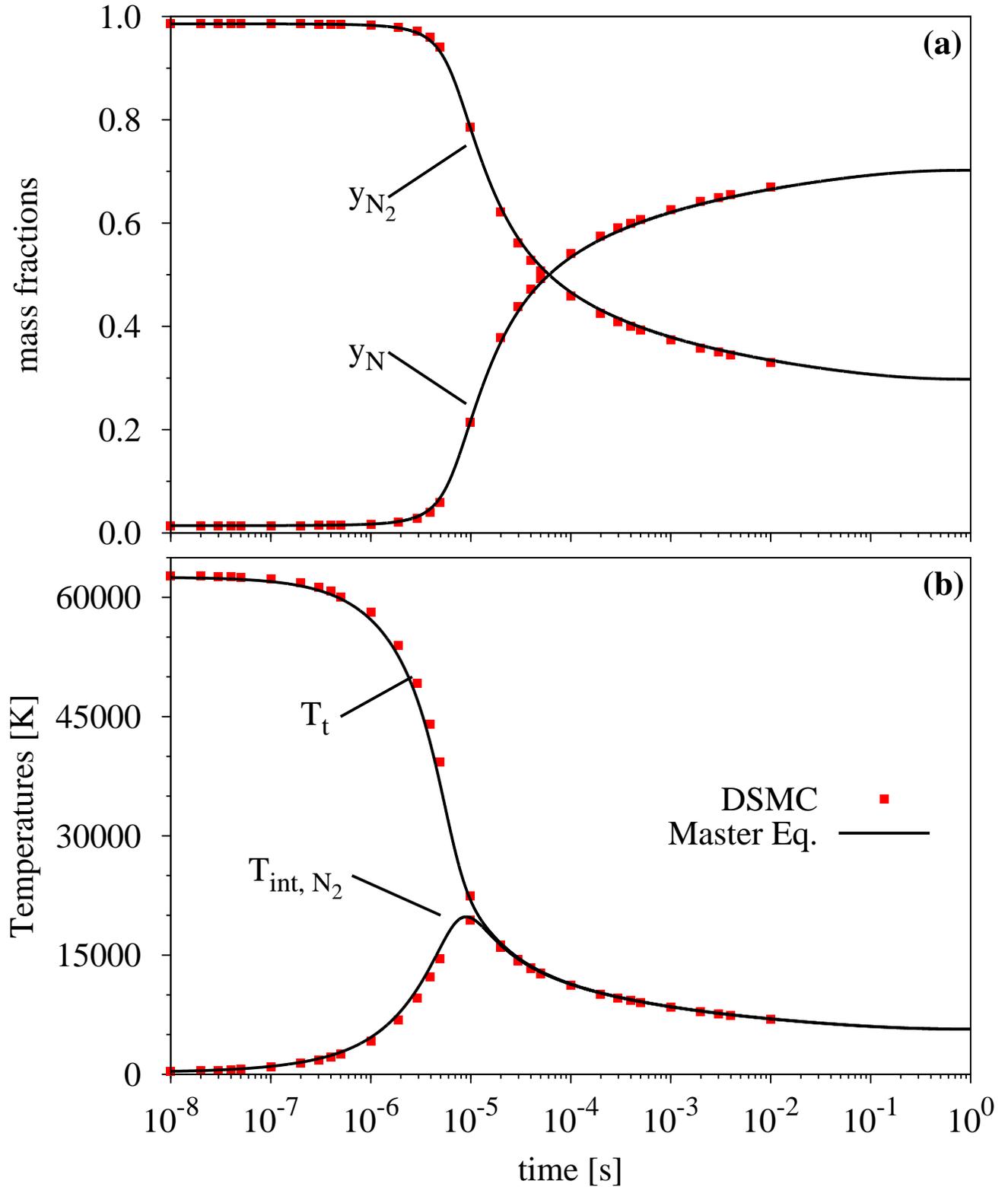}

 \caption{Verification of DSMC implementation using a 100-bin system (with B:P = 90:10 and exponent $n = 2$ in Eq.~(\ref{eq:bin_energy_grid_transformation})) against master equation results. Plot (a): mass fractions of atomic and molecular nitrogen. Plot (b): mixture translational temperature ($T_\mathrm{t}$) and rovibrational mode temperature ($\mathrm{T_\mathrm{int, N_2}}$) of molecular nitrogen. Filled red squares: DSMC flow macroparameters. Solid black lines: corresponding master equation results.}
 \label{fig:verification_variable_bins}
\end{figure}

A comparison of the flow macroparameters extracted from the DSMC simulations and the equivalent master equation profiles is shown in Fig.~\ref{fig:verification_variable_bins}. In this comparison, the 100-bin system with variably-spaced bins (B:P = 90:10, $n = 2$) is used. Fig.~\ref{fig:verification_variable_bins}(a) shows the time evolution of the mass fractions of $\mathrm{N}$ and $\mathrm{N_2}$, while Fig.~\ref{fig:verification_variable_bins}(b) shows the corresponding mixture translational temperature and the internal mode temperature of molecular nitrogen. As can be seen, the DSMC simulations closely match those of the master equations over the first $10^{-2} \, \mathrm{s}$ of the process. Recall that the internal temperature of $\mathrm{N_2}$ is determined using the instantaneous bin populations in the cell, and is an implicit solution to Eq.~(5) in Magin et al.~\cite{magin12a}.


The temperature- and mass fraction curves reveal a slight mismatch between the DSMC- and master equation results. One can see that until approximately $10^{-5} \, \mathrm{s}$, the DSMC profiles lag behind the master equation curves by a small amount. After this point, the relation is inverted, and the DSMC profiles begin to lead the master equation curves. Although the same detailed chemistry rate data and the exact same initial conditions were used in both cases, there still exist differences inherent to the simulation methods themselves. In the hydrodynamic description of the system of master equations, the chemical production terms for individual bin populations and atomic nitrogen are evaluated at the mixture translational temperature $T_{t}$ (See appendix A in Magin et al.~\cite{magin12a}). Furthermore, the rate coefficients obtained from the NASA Ames database were computed as thermal averages of the respective cross sections at equilibrium temperatures $T$. When using these rate coefficients in a simulation, one implicitly assumes that the local instantaneous velocity distributions of all $\mathrm{N_2}$-molecules and $\mathrm{N}$-atoms are nearly Maxwellian at a common temperature $T = T_{t}$, an assumption that is only valid as long as the characteristic time scale for inelastic and reactive processes is significantly larger than the mean collisional time. By contrast, in the kinetic description at the basis of the DSMC method, the molecular velocity distributions may deviate arbitrarily from the Maxwellian equilibrium form, and inelastic processes may take place at similar time scales as the mean collisional time. 

\begin{figure}
 \centering
 \includegraphics[width=\columnwidth]{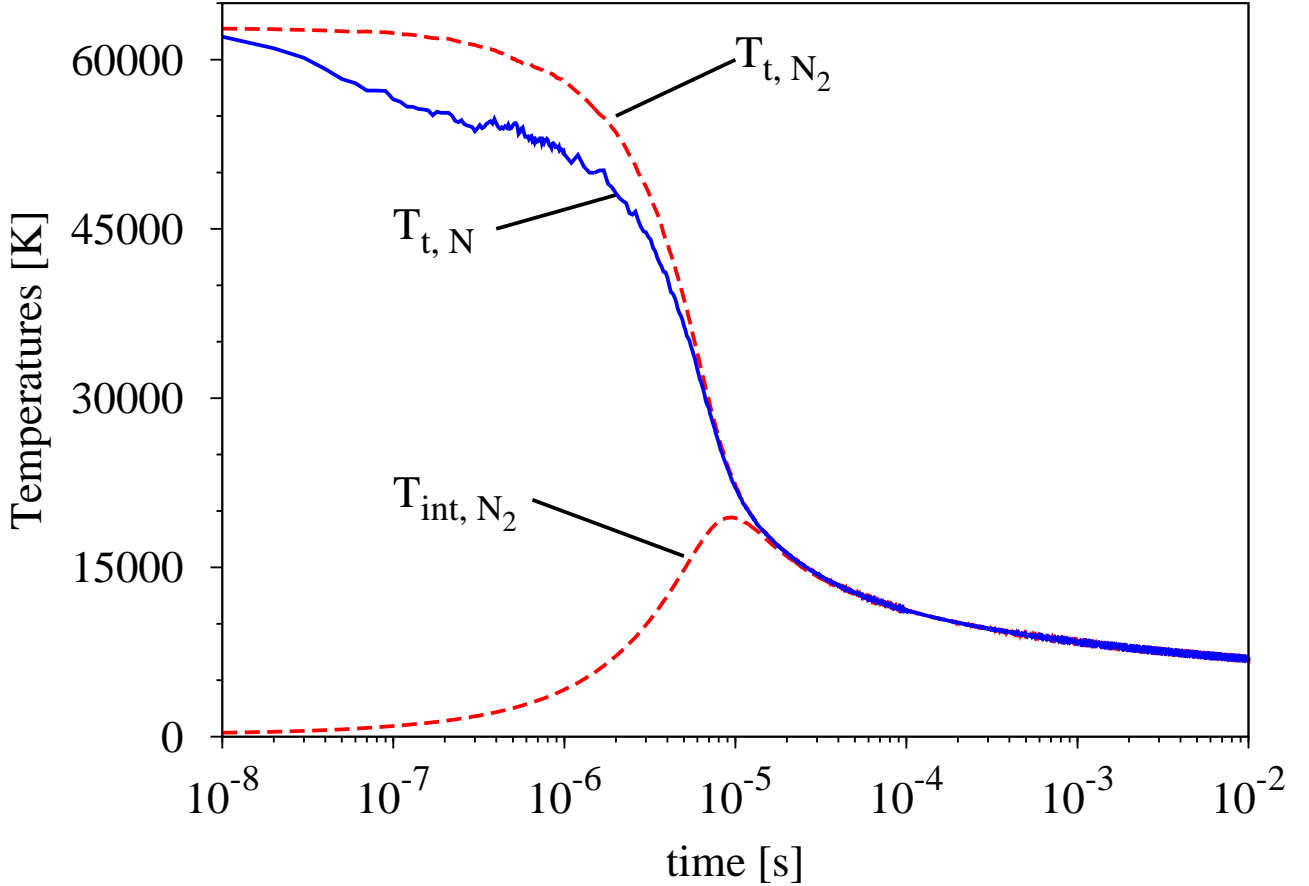}
 \caption{DSMC results with a 100-bin system (with B:P = 90:10 and exponent $n = 2$ in Eq.~(\ref{eq:bin_energy_grid_transformation})). Blue continuous line: translational temperature of atomic nitrogen extracted from DSMC. Red dashed lines: translational ($T_\mathrm{t, N_2}$) and rovibrational mode temperatures ($\mathrm{T_\mathrm{int, N_2}}$) of molecular nitrogen}
 \label{fig:dsmc_species_temperatures}
\end{figure}

Some evidence of this occurring in the initial phases of the reservoir simulations can be seen in Fig.~\ref{fig:dsmc_species_temperatures}. Here the individual translational temperatures of $\mathrm{N_2}$ and $\mathrm{N}$, extracted from the DSMC simulation, are plotted in addition to the internal temperature of $\mathrm{N_2}$. One can see that $T_{t,\mathrm{N}}$ decreases almost immediately after the beginning of the simulation at a much faster pace than $T_{t, \mathrm{N_2}}$. It is only after approximately $10^{-5} \, \mathrm{s}$ that both species temperatures equalize, at about the same time that internal energy relaxation is complete. This initial difference in the species velocity distributions may be responsible for the slightly slower rate of internal energy excitation observed in the DSMC results. Given that in the master equation system all reaction rate coefficients are evaluated at the mixture translational temperature $T_t$, which is much closer to $T_{t, \mathrm{N_2}}$, it is likely that the rates of internal energy exchange at the early stages of the process are over-estimated relative to the DSMC simulation. Incidentally, the $T_{t, \mathrm{N}}$-signal in Fig.~\ref{fig:dsmc_species_temperatures} exhibits significant stochastic noise, due to the relatively small initial number of DSMC particles representing atomic nitrogen. This noise decreases gradually, as more $\mathrm{N}$-atoms are produced. Towards the later stages of the simulation, the stochastic noise level increases again. However, now the noise is caused by the vanishing number of $\mathrm{N_2}$-molecules, as more and more of this species dissociates.


\begin{figure}
 \centering
 \includegraphics[width=\columnwidth]{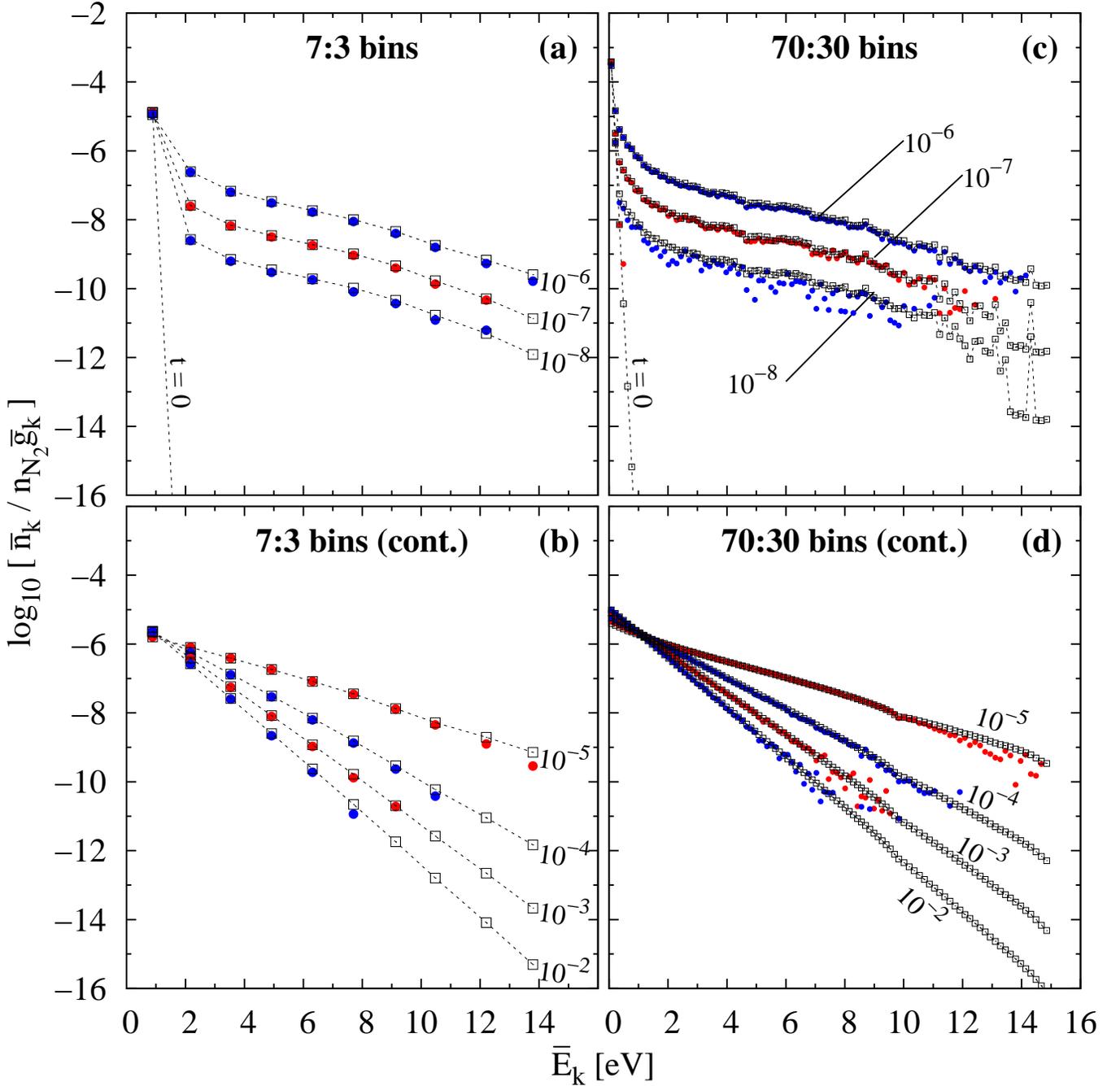}
 \caption{Time evolution of $\mathrm{N_2}$-bin populations for \emph{equally-spaced} bins. Left column shows 7:3-bin system. Plot (a): early stages from $t = 0$ up to $t = 10^{-6} \, \mathrm{s}$, remaining curves for $t = 10^{-5}$ to $t = 10^{-2} \, \mathrm{s}$ shown in plot (b). Plots (c) and (d): equivalent results for the 70:30-bin system. Alternating red and blue circles: bin populations obtained from DSMC at consecutive time steps. Black squares joined by dashed lines: equivalent master equation populations.}
 \label{fig:bin_populations_combined_old_bins}
\end{figure}

In addition to the flow macroparameters, we can gather insight into the time evolution of individual bin populations. They have been extracted from the DSMC- and corresponding master equation results at specific instants during the relaxation-dissociation process. In Figs.~\ref{fig:bin_populations_combined_old_bins}(a) and (b) we first show Boltzmann plots for the 10-bin system using the equally-spaced bins (i.e. B:P=7:3). Then, in Figs.~\ref{fig:bin_populations_combined_old_bins}(c) and (d) we show the corresponding populations using the larger 100-bin (i.e. B:P=70:30) system. The DSMC populations at different time steps are plotted using color-filled symbols, alternating between red squares and blue circles. The equivalent master equation populations are shown as unfilled, black squares, joined by dotted lines for easier visualization. Every such curve is labeled with the time at which the bin populations have been extracted.

The upper row of plots shows the early stages of the process, from $t = 0$, up to $t = 10^{-6} \, \mathrm{s}$, which corresponds to the first 100 DSMC time steps. As can be seen in Fig.~\ref{fig:bin_populations_combined_old_bins}(a), before even the first time step is taken, practically all DSMC particles representing $\mathrm{N_2}$ in the 7:3-bin system populate only the first bin, while the other 9 are empty. This is a consequence of the small number of DSMC particles employed in this simulation, together with the relatively high average energy of the second bin. The value of $\bar{E}_2 = 2.17 \, \mathrm{eV}$ implies an extremely low equilibrium population for the second bin at $T_\mathrm{int}^0 = 300 \, \mathrm{K}$. Since in the current simulations at most about $380\,000$ DSMC particles are being used to populate all of the rovibrational bins of $\mathrm{N_2}$ (and this particle number steadily decreases as dissociation progresses), it is impossible to accurately account for the bin populations below a certain threshold. This can be observed wherever there is a mismatch between the bin populations predicted by DSMC and the master equations. The DSMC populations become ``noisy'', or drop off significantly compared to the master equation values. In fact, the correct linear slope of the initial Boltzmann distribution at $T_\mathrm{int}^0 = 300 \, \mathrm{K}$, can only be inferred from the dotted line of the master equation curve. Still in Fig.~\ref{fig:bin_populations_combined_old_bins}(a), and after only a few time steps, the higher-lying bins gradually become populated. One can see that the population rise in the higher-lying bins (2-10) proceeds fairly evenly, preserving the general shape of the high-energy part of the distribution. Then, excitation of the rovibrational modes continues up to approximately $10^{-6} \, \mathrm{s}$. This corresponds to 1000 DSMC time steps, and incidentally is the moment at which $T_\mathrm{int}$ reaches its maximum. The populations from this time on are shown in Fig.~\ref{fig:bin_populations_combined_old_bins}(b). As time advances further, the bins in the 7:3-bin system now seem to be populated according to near-Boltzmann distributions with slowly increasing negative slopes. Thus, at this stage the rovibrational mode has almost reached equilibrium at the slowly decreasing $T_\mathrm{int}$. The resolution limit of the DSMC results becomes noticeable for times $10^{-5} \, \mathrm{s}$ and later: in bins where $\log_{10} \left[ \bar{n}_k / n_\mathrm{N_2} \bar{g}_k \right] < -11$, practically no DSMC particles are recorded. This mostly affects the high-energy tail of the distribution. Fortunately, these bins seem to contribute little to the overall behavior of the flow macroparameters. Thus, despite the mismatch in bin populations, general agreement between the DSMC- and master equation results remains very good.

The higher resolution of the 100-bin system now allows us to take a more detailed look at the bin populations. Fig.~\ref{fig:bin_populations_combined_old_bins}(c) shows the results at the early stages of the relaxation for the 70:30-bin case. Compared to the 7:3-bin case, the Boltzmann distribution at $T_\mathrm{int}^0 = 300 \, \mathrm{K}$ is now represented by several low-energy bins before the populations drop below the accuracy limit of $380\,000$ DSMC particles. From $t = 10^{-8}$ to $10^{-6} \, \mathrm{s}$ we can now observe distinct non-equilibrium distributions, with the higher-lying bins slowly becoming populated over time. Similar to the 7:3-bin case, the excitation of these higher-lying bins seems to proceed fairly uniformly across the internal energy scale. Over the first 100 time steps, one can still roughly distinguish the steep linear decay of the initial low-temperature Boltzmann distribution from the much flatter slope dominating the high-energy tail. Given the ten times higher resolution, the transition between the two behaviors is smoother than in the 7:3-bin case. In the master equation results small ``wiggles'' can now be spotted. Compared to the 7:3-bin case, these population curves appear less smooth, and the oscillations seem to increase towards the high-energy tail, beyond $\Delta E_{(v=0, J=0)}^{D} = 9.75 \, \mathrm{eV}$ (especially at $10^{-8}$ and $10^{-7} \, \mathrm{s}$). Unlike DSMC, the master equation results are not subject to particle-related stochastic noise, and we must conclude that these oscillations are a feature of the underlying N3-reaction mechanism. In Fig.~\ref{fig:bin_populations_combined_old_bins}(a), it was merely the lower resolution of the 7:3-bin bin model that prevented us from distinguishing these features. In addition to this, both in Fig.~\ref{fig:bin_populations_combined_old_bins}(c) and (d) the noise in the DSMC results now becomes apparent at levels of approximately $\log_{10} \left[  \bar{n}_k / n_\mathrm{N_2} \bar{g}_k\right] < -8$. This decreased accuracy in representing the sparsely populated bins is consistent with the fact that compared to the 7:3-bin case, there are now ten times as many bins to be filled by the same amount of particles. 

For times $10^{-5}\, \mathrm{s}$ and later, the DSMC populations at the high-energy tail become too noisy to provide any useful information. In Fig.~\ref{fig:bin_populations_combined_old_bins}(d) we must rely on the master equation results, if we wish to distinguish any features in bins with energies  above $7 \, \mathrm{eV}$. As was predicted in the 7:3-bin case, the bin populations are now much closer to Boltzmann distributions, and their slopes gradually become steeper over time, indicating a decrease in the internal temperature. Upon closer inspection though, a small ``kink'' in the distribution slopes can be observed, exactly at $\Delta E_{(v=0, J=0)}^{D} = 9.75 \, \mathrm{eV}$, the limiting energy separating bound- from predissociated bins. We see this effect only in the master equation results, since the populations in question are well below the threshold accessible to our DSMC simulations. The good agreement between the two methods at the macroparameter scale makes us conclude that this subtle feature is of secondary importance in the overall behavior of the N3-reaction mechanism.

\begin{figure}
 \centering
 \includegraphics[width=\columnwidth]{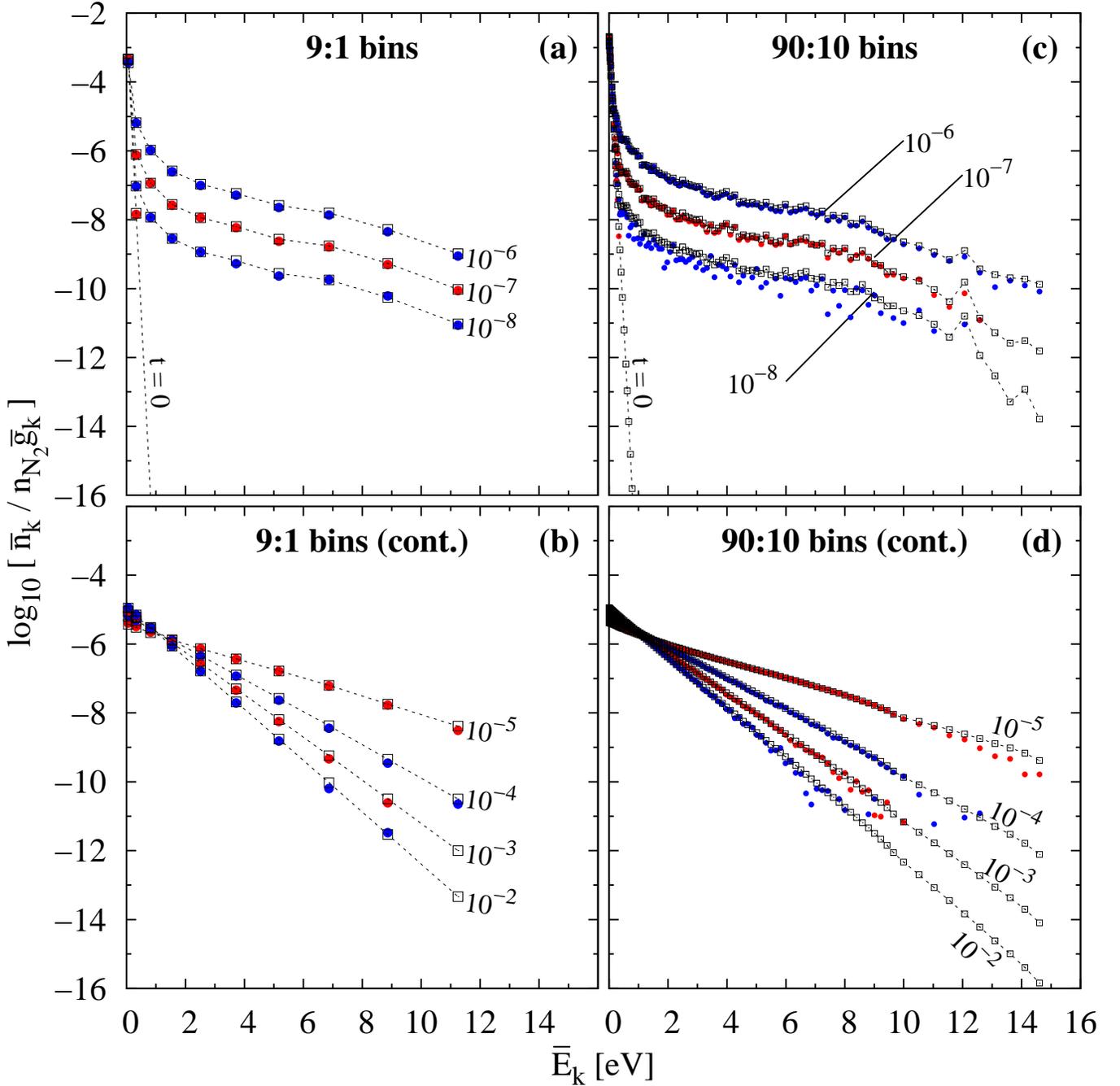}
 \caption{Time evolution of $\mathrm{N_2}$-bin populations for \emph{variably-spaced} bins. Left column shows 9:1-bin system. Plot (a): early stages from $t = 0$ up to $t = 10^{-6} \, \mathrm{s}$, remaining curves for $t = 10^{-5}$ to $t = 10^{-2} \, \mathrm{s}$ shown in plot (b). Plots (c) and (d): equivalent results for the 90:10-bin system. Alternating red and blue circles: bin populations obtained from DSMC at consecutive time steps. Black squares joined by dashed lines: equivalent master equation populations.}
 \label{fig:bin_populations_combined_new_bins}
\end{figure}

Now we present the corresponding plots for the variably-spaced bins. The 10-bin case is shown in Figs.~\ref{fig:bin_populations_combined_new_bins}(a) and (b), whereas the 100-bin case is shown in Figs.~\ref{fig:bin_populations_combined_new_bins}(c) and (d). All bin populations were extracted at the same time steps as for Fig.~\ref{fig:bin_populations_combined_old_bins}. Focusing on the 9:1-bin case first, we can see that, compared to the 7:3-bin case of Fig.~\ref{fig:bin_populations_combined_old_bins}(a), all population symbols have shifted to the left. This is due to the re-definition of the bins discussed in Sec.~\ref{sec:variably_sized_bins}. The bin populations are plotted on the abscissa at the location of the bin-average energies, which now have shifted toward lower values. Notice that thanks to this adjustment one is able to better resolve populations in the lower-energy region. Compared to the equally-spaced bins, this makes it possible to reproduce the steeply decaying Boltzmann distribution at $T_\mathrm{int} = 300  \mathrm{K}$ with greater accuracy. Even with a fairly small number of DSMC particles, the populations of the three lowest-energy bins can now be accurately represented.

Over the course of the first 100 DSMC time steps, i.e. for times $t = 10^{-8} - 10^{-6} \, \mathrm{s}$,  we again see the bins slowly adopting a distribution dominated by two separate behaviors. The lower-energy bins roughly retain the shape of the initial Boltzmann distribution, whereas the higher-energy bins are populated according to a separate, much ``flatter'' distribution. Compared to the rather abrupt change in slope between the curves joining bins 1-2 and bins 2-10 respectively, as previously seen in Fig.~\ref{fig:bin_populations_combined_old_bins}(a), with the variably-spaced bins of Fig.~\ref{fig:bin_populations_combined_new_bins}(a) we are now able to observe a much smoother transition between the two behaviors. Recall also that the highest-energy bin in this 9:1-bin system covers the range of all pre-dissociated levels, whereas the same region was represented by three separate bins in the original 7:3-bin system. However, this has not substantially altered the slopes of the distributions at the high-energy end. In fact, one slight advantage of this arrangement is that a single bin now collects all DSMC particles with internal energies above the dissociation limit, instead of having them spread out over three bins. As a consequence, the populations of the higher-energy bins can be determined with greater accuracy than in the original arrangement. We can observe this by comparing the populations extracted from DSMC with those from the master equations. For example, whereas in Fig.~\ref{fig:bin_populations_combined_old_bins}(b), at times $t = 10^{-5} - 10^{-2} \, \mathrm{s}$, the DSMC-extracted populations of the three predissociated bins are mostly empty, with the equivalent ones of Fig.~\ref{fig:bin_populations_combined_new_bins}(b) this problem is confined to the highest-energy bin.

The re-arrangement of bin energies is also reflected in the populations of the 90:10-bin system, shown in Figs.~\ref{fig:bin_populations_combined_new_bins}(c) and (d). The main difference with respect to the 70:30-bin system is that the pre-dissociated levels are now represented by only 10 bins. This affects the small oscillations in bin populations above $\Delta E_{(v=0, J=0)}^D$, which we had noticed previously in Fig.~\ref{fig:bin_populations_combined_old_bins}(c). Similar wiggles appear now at times $t = 10^{-8} - 10^{-6} \, \mathrm{s}$ in Fig.~\ref{fig:bin_populations_combined_new_bins}(c), but the shapes of the distributions are different. As was the case before, since these bins lie below the sensitivity threshold of our DSMC simulations, we can only observe this effect with the help of the master equation results. The small ``kink'' seen previously in Fig.~\ref{fig:bin_populations_combined_old_bins}(d) affecting the populations at the later stages is still visible in Fig.~\ref{fig:bin_populations_combined_new_bins}(d) with the 90:10-bin arrangement. But again, it is only revealed in the master equation results. There are still too few DSMC particles populating the predissociated bins to capture this subtle effect.

\begin{figure}
 \centering
 \includegraphics[width=\columnwidth]{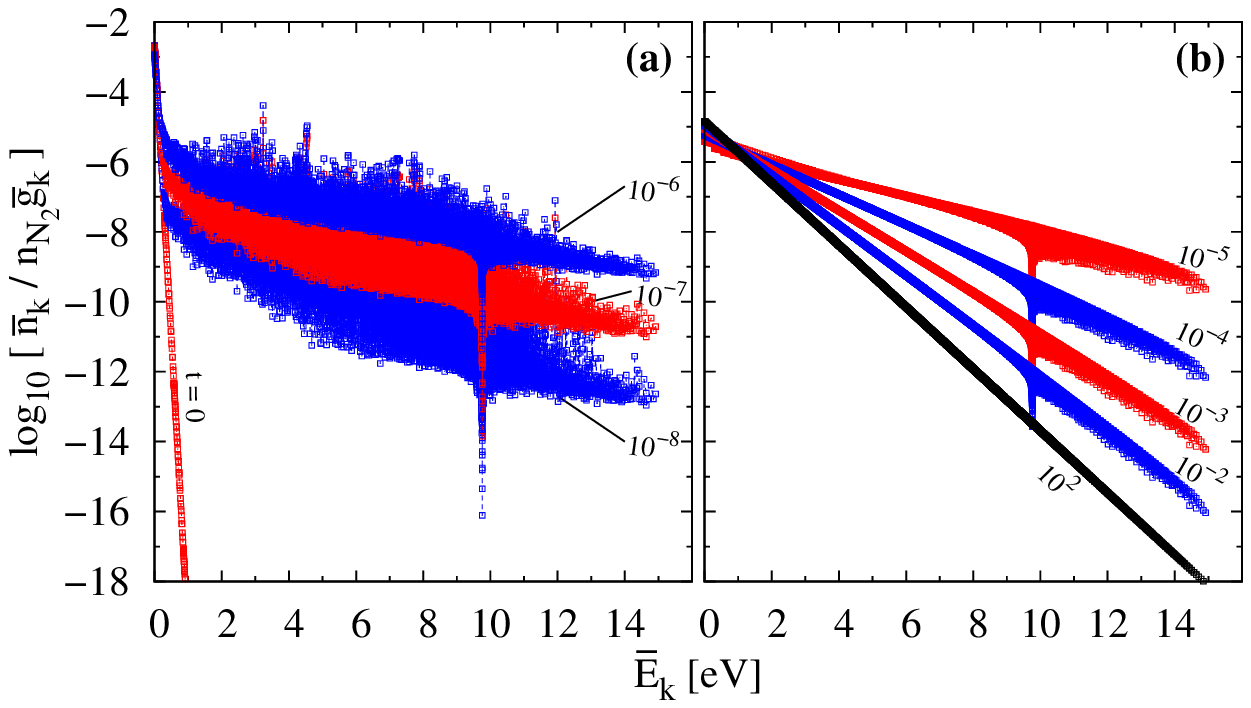}
 \caption{Rovibrational populations extracted from master equation calculation with the full 9390 levels. Successive time steps plotted as alternating red and blue squares joined by dotted lines. Plot (a): instantaneous populations from $t = 0$ to $t = 10^{-6} \, \mathrm{s}$. Plot (b): remaining times from $t = 10^{-6}$ to $t = 10^{2} \, \mathrm{s}$. Final populations at $t = 10^{2} \, \mathrm{s}$ (black squares) follow a Boltzmann distribution at the system's equilibrium temperature.}
 \label{fig:full_level_populations_combined}
\end{figure}

Although unrelated to the verification of our DSMC code, the ``noise'' in the master equation results merits a closer look. We now examine Boltzmann plots of individual level populations, which we extracted from the master equation calculations using the full detailed chemistry mechanism, with all 9390 rovibrational levels. Recall from Table~\ref{tab:case_2_reservoir_conditions} that these calculations had been used as the reference results in all comparisons. We extracted these level populations for the same times during the simulation as shown in Figs.~\ref{fig:bin_populations_combined_old_bins} and \ref{fig:bin_populations_combined_new_bins}. Even though it was found in Sec.~\ref{sec:parametric_study_kinetics} that the detailed chemistry with the full set of levels behaves somewhat differently from the binned systems (faster initial relaxation, but slower dissociation later on, see Figs.~\ref{fig:master_equation_results_old_bins} and \ref{fig:master_equation_results_new_bins}), the results at the given time steps are still close enough to make a qualitative comparison. The level populations are now shown in Figs.~\ref{fig:full_level_populations_combined}(a) and (b). As can be seen, except for the clear straight-line Boltzmann distributions at $t = 0 \, \mathrm{s}$, and the final one at $t=10^{2} \, \mathrm{s}$, the shape of the distributions is much more complex than in the binned systems. From the beginning on, we see in Fig.~\ref{fig:full_level_populations_combined}(a) that, despite there being a general trend for the overall distribution at a given time, the populations from neighboring rovibrational energy levels tend to vary by many orders of magnitude in a seemingly random manner. The vertical spread becomes less severe in Fig.~\ref{fig:full_level_populations_combined}(b), where is seems that the level populations slowly converge toward a common distribution, at a single internal temperature. However, large jumps in population are still visible for much of the remaining time, before final equilibrium is reached. The major common feature in all curves is the precipitous drop in populations affecting a few levels closest to the dissociation energy. This is true for both the bound levels to the left, as well as the predissociated levels on the right of $\Delta E_{(v=0, J=0)}^D = 9.75 \, \mathrm{eV}$. It seems that the probability of dissociation is especially high for molecules populating these levels, which is why they remain depleted during the entire process.

As can be seen from comparing Figs.~\ref{fig:bin_populations_combined_old_bins}-\ref{fig:full_level_populations_combined}, the binning procedure ``filters out'' a lot of the fine details of the full chemistry mechanism, regardless of whether the original, or the alternative bins are used. Some indication of the underlying fine structure is revealed as the number of bins is increased, but even with as many as 100 bins, most of the details of the full level populations remains smoothed out. One must accept this loss of detail with the bin model, and this might become a problem if one is interested in accurately determining the exact populations of individual rovibrational levels at any given instant. However, as long as one is only interested in the overall evolution of the gas in terms of energy content and species concentrations, the bin models should provide enough accuracy.

It might come as somewhat of a surprise that the populations extracted from DSMC and the master equations match reasonably well at nearly all time steps examined. Due to the inherent differences between the kinetic- and hydrodynamic-scale approaches, one might have expected greater discrepancies, especially at the early stages of the process, before the rovibrational mode has thermalized to its own temperature. However, the good agreement between the two approaches confirms, at least for these space-homogeneous calculations, that the two approaches are roughly equivalent. This gives us confidence in the verification of the DSMC collision routine implementation. Obviously, there will always remain a practical limit to how well we could resolve the sparsely-populated bins with limited number of particles per cell.


\subsection{Computational cost of DSMC implementation} \label{sec:dsmc_computational_cost}

It is worthwhile to briefly discuss the computational cost associated with our DSMC implementation. The comparison presented here is not rigorous in an absolute sense, since run-times depend on the details of the test case, the type of processor used and on the manner in which the collision algorithms are implemented. However, comparing the relative run-times listed in Table~\ref{tab:case_2_run_times} can give an indication of the computational cost as a function of the number of bins. Since in the present test case the particle motion routines of the DSMC code are inactive, practically all the effort is spent on the collision routines. There are two main parts of the algorithm responsible for the run-time: 1) computation of the bin-specific total cross section and 2) determination of the collision outcome. Both these parts require a loop over all bins, which makes the computation time directly proportional to the overall number of discrete internal states. This is confirmed by the values listed in Table~\ref{tab:case_2_run_times}. 
\begin{table}
 \centering
 \caption{Run times for DSMC code ($10^6$ time steps) using Intel\textsuperscript{\textregistered} Xeon\textsuperscript{\textregistered} X5650 CPU at 2.67GHz for variable-sized bins}
 \label{tab:case_2_run_times}

 \begin{tabular}{r | c}
           & CPU time [h] \\ \hline
   10 bins & 16.4 \\
   20 bins & 24.5 \\
  100 bins & 93.1 \\
  200 bins & 200 \\
 \end{tabular}
\end{table}

Extrapolating from these numbers, we estimate that running the same test case until time $t = 10^{-2} \, \mathrm{s}$ for the full set of levels, and using the same computational parameters otherwise, would require slightly more than a full year. Thus, it was clearly impractical to use the full set of levels in our DSMC implementation.


\subsection{Effect of neglecting recombination reactions} \label{sec:neglecting_recombination}

\begin{figure}
 \centering
 \includegraphics[width=\columnwidth]{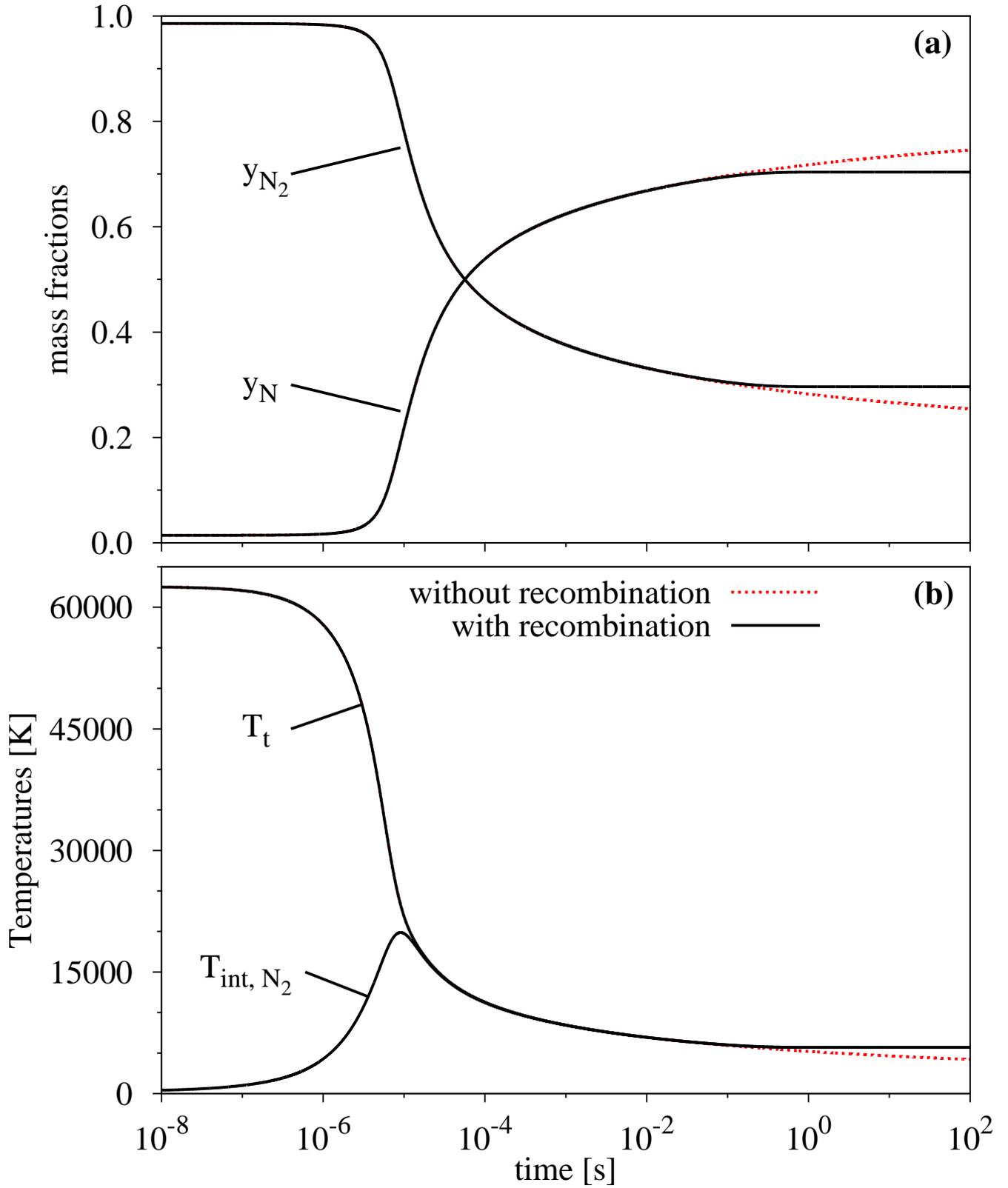}
 \caption{Master equation calculations for a 200-equally-spaced-bin system (B:P = 140:60, $n = 1$), including recombination reactions (solid black lines) vs. neglecting them (red dotted lines). Plot (a): mass fraction profiles for molecular and atomic nitrogen, plot (b): mixture translational temperature and rovibrational mode temperature of $\mathrm{N_2}$.}
 \label{fig:recombination_comparison}
\end{figure}

Notice that in the DSMC macroparameter profiles of Figs.~\ref{fig:verification_variable_bins} and \ref{fig:dsmc_species_temperatures} all DSMC results are shown only up to time $t = 10^{-2} \, \mathrm{s}$, cutting off the final stages of the approach towards equilibrium. There are two reasons for this. First, using the above-mentioned constant $\Delta t = 10^{-8} \, \mathrm{s}$ in the DSMC simulations would require a total of about $10^{8}$ time steps to reach $t = 10^{0} \, \mathrm{s}$, approximately the physical time needed to reach equilibrium. With the present code, such a simulation would require several weeks, or months to be completed. Since the goal of this test case was to verify the correct DSMC implementation of the bin model and its ability to reproduce the master equation results for the early stages of the relaxation process, we considered it enough to run the DSMC simulations until reaching $t = 0.01$ seconds. By comparison, the implicit time integration scheme used in the master equation solver is able to adapt the computational time step and provides the full solution in less than one second of CPU time. The second reason for stopping the DSMC runs at $t = 0.01 \, \mathrm{s}$ is that in the current DSMC implementation recombination reactions, i.e. $\mathrm{3 \, N} \rightarrow \mathrm{N_2 \left(k\right) + N}$, are neglected. This makes it impossible for the DSMC results to match the final equilibrium concentrations of $\mathrm{N}$ and $\mathrm{N_2}$, as predicted by the master equations. Neglecting recombination reactions is not a problem for the early stages of the process, when the temperature in the reservoir is high and dissociation of molecular nitrogen is the dominant reaction. However, once the gas has cooled down enough, the recombination rate should gradually increase to balance out the declining dissociation rate, finally reaching dynamic equilibrium in the reservoir. Since recombination has been neglected in the DSMC implementation, it is incapable of reproducing this equilibrium regime. For illustration purposes the difference in behavior is demonstrated in Fig.~\ref{fig:recombination_comparison}, where master equation results for the original 140:60-bin system are shown with- and without recombination enabled. When recombination reactions are not taken into account (red dotted lines), the remaining $\mathrm{N_2}$ will eventually dissociate completely, and the final state of the system diverges from the correct equilibrium state. On the contrary, if recombination is included, the system reaches the final equilibrium state predicted in Table~\ref{tab:case_2_reservoir_conditions} at approximately $10^0 \, \mathrm{s}$ (solid black lines).

\section{Conclusions} \label{sec:conclusions}

In this paper we have presented the integration of an efficient coarse-grain rovibrational model for internal energy excitation and molecular dissociation in nitrogen with the Direct Simulation Monte Carlo method. The state-resolved cross sections used in our chemical reaction mechanism were generated using a method outlined in an accompanying publication~\cite{torres18a}. We have verified our DSMC implementation against master equation results obtained using state-resolved reaction rates, directly derived from our own cross sections.

As a first step, within the constraints imposed by the energy-based uniform rovibrational (URVC) bin model, we have proposed a new, alternative approach to lumping together the rovibrational energy levels of $\mathrm{N_2}$. In contrast to the original approach, where all bins were defined using \emph{equally-spaced} energy intervals, we have opted for \emph{variably-spaced} energy bins. This means we adapted the size of each individual bin to improve agreement of the gas thermodynamic properties predicted by the coarse-grained system with those of the full set of rovibrational levels, while maintaining the overall number of bins equal to the original formulation. This re-definition of the bin boundaries emphasizes a higher resolution at the low-energy limit of the rovibrational energy scale, by placing more bins containing fewer levels in this region. We found the improvement to be significant, because this re-distribution allows us to closely match the thermodynamic properties of the full system (which are based on Boltzmann-weighted populations, and thus mostly influenced by the lowest-lying energy levels) by using as few as 10 variably-spaced bins. 

However, we are aware that clustering more bins at the low-energy limit comes at the expense of accurately resolving the higher-energy level populations. If taken too far, this can cause problems, since in chemically reacting flows the high-energy level populations are known to significantly affect the dissociation rates. In order to assess the influence of our new binning approach on the chemical-kinetic behavior of the overall gas mixture, we performed a series of heat bath calculations using a master equation solver. We simulated internal energy excitation and dissociation of a suddenly heated gas in an adiabatic reactor for three different initial translational temperatures (approximately $30\,000$, $60\,000$ and $120\,000$ kelvin). We compared the results obtained with both binning approaches to reference calculations using the full set of rovibrational levels. In the low- and medium temperature cases the new binning approach showed much closer agreement with the full system. This improvement is especially noticeable when the total number of bins is small. At the highest temperature investigated, the results are less conclusive. Both binning strategies struggle to match the behavior of the full system when only a small number of bins is used. Furthermore, convergence to the reference solution is not even achieved with as many as 1000 bins. 

This partial failure suggests that there are limitations to the uniform binning approach, which that cannot be overcome simply by re-shuffling the rovibrational levels among bins, and that our coarse-grain model can only be expected to yield accurate results within a certain range of flow conditions. On the other hand, one should consider that the highest-temperature conditions investigated in our study are far beyond what one would typically encounter in atmospheric entry flows. At such extreme conditions, the precise rates of rovibrational excitation and dissociation of molecular species would be of secondary importance, given that the gas would almost immediately dissociate, and most certainly be electronically excited and ionized. We should also note that, since in the master equation studies we used state-resolved reaction rate coefficients fully consistent with the cross sections of our DSMC implementation, the considerations just presented remain valid, regardless of whether a hydrodynamic or gas-kinetic description is used. Thus, and our alternative binning can be applied both in classical CFD and in DSMC codes.


In the second part of this paper we discussed details related to the integration of our coarse-grain rovibrational model with the DSMC method. We outlined the major adaptations to implement the state-to-state chemistry mechanism for excitation and dissociation in $\mathrm{N_2}$ in our purpose-built DSMC code. The discussion was limited to diatom-atom (in this case $\mathrm{N}$+$\mathrm{N_2}(k)$) collisions, given the detailed cross section data available to us. However, there exists no fundamental restriction to extending the algorithms to $\mathrm{N_2}$-$\mathrm{N_2}$ collisions, or interactions between other chemical species. The only requirement is that state-resolved reaction cross sections be available for all colliding species.

Although we re-used as much of the machinery of standard DSMC as needed, we disregarded some of the conventions, considered unsuitable in our modeling of inelastic collisions. One of them was the use of a single total cross section based on the VHS model for all $\mathrm{N_2}$-$\mathrm{N}$ encounters. Instead, we compute an internal-state-specific total cross section $\sigma_{k}^T \left( E_t \right)$ for each particular pre-collision bin, based on our state-resolved cross sections. Even though our cross sections are expressed in terms of simple analytical functions, they contain internal state-specific tuning parameters emanating from the \emph{ab initio} Ames N3 database. Special care was taken to compute the probabilities of post-collision outcomes to precisely reproduce the expected rates of internal energy excitation and dissociation observed in the hydrodynamic description. This meant that constraints relating the forward and backward excitation cross sections had to be enforced, in order to ensure micro-reversibility.

We were able verify our DSMC implementation by comparing to master equation results from the previously used test case simulating strong thermo-chemical nonequilibrium in a suddenly heated gas. In addition to the macroscopic moments, i.e. species concentrations and kinetic- and internal mode temperatures, we also compared the detailed bin populations extracted from DSMC with equivalent quantities from the master equation results. With DSMC our ability to accurately resolve the bin populations is limited to the most populated bins, because those will contain the majority of DSMC simulator particles. Using the master equation results as a reference, we can observe where the stochastic noise of DSMC starts to overwhelm the signal in the lower-populated internal states.

One important limitation of our current DSMC implementation is that we disregard molecular recombination reactions. At the lower-temperature conditions, this makes it it impossible for us to attain the expected equilibrium composition in our adiabatic reactor simulations, and in general limits our ability to accurately model the chemistry of cooling and expanding flows. In our view, a fully consistent treatment of recombination in a state-to-state description DSMC remains an open problem, and no satisfactory answer has so far been proposed in literature.

We conducted a non-exhaustive benchmarking of the DSMC implementation. Computational time per DSMC time step was found to be roughly proportional to the number of energy bins. This is to be expected, given the need to compute the total cross sections and selecting a post-collision outcome for each $\mathrm{N_2}(k)$+$\mathrm{N}$ collision. However, we expect that when including $\mathrm{N_2} (k_1)$+$\mathrm{N_2}(k_2)$ collisions in the future, the computational effort would become proportional to the square of $n_\mathrm{bins}$, and this is a strong reason to for using as few bins as possible.

One should also not forget that in the present DSMC calculations no spatial flow field gradients were present. Thus, transport phenomena related to viscosity, species diffusion, or heat conduction were of no concern here. This was fortunate, because it allowed us to ignore for the time being the uncertainty in our elastic cross sections (see the discussion in our accompanying paper~\cite{torres18a}). This issue can no longer be ignored, once we move to simulating actual hypersonic flows in 1D, or higher spatial dimensions. The thickness and internal structure of the bow shock ahead of a blunt-nosed object immersed in a hypersonic flow will be affected by the magnitude of the elastic, vs. the inelastic and/or reactive cross sections. The same is to be expected for the flow field in the stagnation point region. However, addressing these issues is beyond the scope of the current paper. In future work we plan to examine these issues more closely and use our DSMC model to study atmospheric-entry-type flow configurations. There we will be able to resolve the shock structure and to observe the incubation length for dissociation.


\section*{Acknowledgments}

We wish to thank Dr. A. Munaf\`o for his help with the master equation solver used in our parametric studies and the verification of our DSMC code. We also wish to thank Dr. R.L. Jaffe for his useful comments regarding the NASA Ames database. Furthermore, we would like to thank Prof. T. Schwartzentruber for his help and suggestions in writing the original DSMC code used in this comparison.

Research of T.E. Magin was sponsored by the European Research Council Starting Grant \#259354.


\appendix

\section{Determination of initial conditions for reservoir calculations} \label{sec:initial_conditions_explanation}

The initial conditions for the adiabatic heat bath simulations discussed in Sec.~\ref{sec:parametric_study_kinetics} were determined in the following manner. We first assumed a supersonic flow of nitrogen at free-stream temperature $T_\infty = 300 \, \mathrm{K}$, pressure $p_\infty = 5.0413 \, \mathrm{m Pa}$ and velocity $u_\infty$ (See the first column of Table.~\ref{tab:case_2_reservoir_conditions} for the velocities chosen). These three values roughly cover the range of velocities at which hypersonic vehicles travel during Earth atmospheric entry. The value of the free-stream pressure indirectly affects our simulation results by fixing the molecular collision rate in the gas reservoir and the associated time scale of the excitation-dissociation process. Since our comparison calculations all use the same set of initial conditions though, all results are affected in an equal manner by this arbitrary choice. Apart from the artificially added atomic nitrogen (mole fraction $x_{\mathrm{N},\infty} = 0.028$) the free-stream gas is assumed to be uniform and in equilibrium. We then determined the initial translational temperature for the heat bath calculations by assuming that the flow kinetic energy is converted entirely into thermal energy of the translational mode. The initial translational temperature and reservoir pressure were then determined as: 

\begin{equation}
 T_{\mathrm{t},0} = \left( 1 + \frac{\gamma_\mathrm{t} - 1}{2} \mathrm{Ma_\infty^2} \right) T_\infty, \label{eq:total_temperature_translation}
\end{equation}

and

\begin{equation}
 p_{0} = p_\infty \left( \frac{T_{\mathrm{t},0}}{T_\infty} \right)^{\gamma_\mathrm{t} / \gamma_\mathrm{t} - 1}
\end{equation}

respectively. Here, $\gamma_\mathrm{t} = 5/3$ is the heat capacity ratio translational energy mode of the free-stream gas. Recall that the internal energy modes of the gas are initially assumed to remain frozen at the free-stream temperature. The Mach number in Eq.~(\ref{eq:total_temperature_translation}) is calculated via:

\begin{equation}
 \mathrm{Ma_\infty} = \frac{u_\infty}{\sqrt{\gamma_\mathrm{t} \, \mathrm{k_B} / \bar{m}_{\mathrm{mix}, \infty} T_\infty}} \label{eq:total_pressure_translation}
\end{equation}

where $u_\infty$ is the free-stream velocity and $\bar{m}_{\mathrm{mix}, \infty} = (1 - x_\mathrm{N, \infty}) m_\mathrm{N_2} + x_\mathrm{N, \infty} m_\mathrm{N}$ is the average molecular mass of the free-stream mixture, taking into account both the molecular and atomic nitrogen initially present.

\section{Mass production terms for macroscopic balance equations in bin model} \label{sec:urvc_bin_model_macroscopic_balance_equations}

The set of master equations used for verification in Sec.~\ref{sec:verification} has the following form:
\begin{align}
 \frac{\mathrm{d} \bar{n}_{k}}{\mathrm{d} t} & = \frac{\bar{\omega}_k}{m_\mathrm{N_2}} \qquad k \in \mathcal{K}_\mathrm{BP} \label{eq:bin_master_equations_n2} \\
 \frac{\mathrm{d} n_\mathrm{N}}{\mathrm{d} t} & = \frac{\omega_\mathrm{N}}{m_\mathrm{N}} \label{eq:bin_master_equation_n}.
\end{align}

Here, the $\bar{n}_k$ are the number densities of nitrogen molecules populating rovibrational bin $k$, and $n_\mathrm{N}$ is the number density of atomic nitrogen. 
The chemical source terms on the right hand side are given by Eq.~(\ref{eq:molecular_nitrogen_bins_source_term}) and (\ref{eq:atomic_nitrogen_reactive_source_term_bins}) respectively. The system is assumed to be isolated from its surroundings, which implies that its mass, momentum and energy remain constant over time. As a consequence, the gas velocity can be arbitrarily set to zero and the energy balance can be re-arranged to yield the instantaneous translational temperature of the mixture, based only on the mixture composition:
\begin{equation}
 T_\mathrm{t} = \frac{\rho e_0 - \left[ \sum_{k \in \mathcal{K}_\mathrm{BP}} \bar{n}_k \bar{E}_k + n_\mathrm{N} E_\mathrm{N} \right]}{\frac{3}{2} \mathrm{k_B} \left[ \sum_{k \in \mathcal{K}_\mathrm{BP}} \bar{n}_k + n_\mathrm{N} \right]},
\end{equation}
with the density $\rho$ remaining constant and the initial energy of the system imposed as:
\begin{equation}
 \begin{split}
  e_0 & = \left( \frac{y_\mathrm{N}^0}{m_\mathrm{N}} + \frac{1 - y_\mathrm{N}^0}{m_\mathrm{N_2}} \right) \frac{3}{2} \, \mathrm{k_B} T_\mathrm{t}^0 + \frac{y_\mathrm{N}^0}{m_\mathrm{N}} E_\mathrm{N} \\
  & + \frac{\left( 1 - y_\mathrm{N}^0 \right)}{m_\mathrm{N_2}} \sum_{k \in \mathcal{K}_\mathrm{BP}} \frac{\bar{g}_k \, \bar{E}_k}{\bar{Q}_\mathrm{N_2}^\mathrm{int} \left( T_\mathrm{int}^0 \right)} \exp \left( \frac{- \bar{E}_k}{\mathrm{k_B} T_\mathrm{int}^0} \right).
 \end{split}
\end{equation}

Here, $T_\mathrm{t}^0$ and $T_\mathrm{int}^0$ are the initial translational and internal temperatures respectively, and $y_\mathrm{N}^0$ the initial mass fraction of atomic nitrogen. The partition function of the internal mode is given by Eq.~(\ref{eq:bin_internal_mode_partition_function}). The mass production term for bin $k$ due to inelastic $\mathrm{N_2}\left( k \right) + \mathrm{N}$-collisions is given by:
\begin{equation}
 \begin{split}
 \frac{\bar{\omega}_k}{m_\mathrm{N_2}} & = \sum_{\substack{l \in \mathcal{K}_{\mathrm{BP}}\\ \left( k < l \right)}} \left\lbrace - \bar{k}_{k \rightarrow l}^E \, \bar{n}_k \, n_\mathrm{N} + \frac{\bar{k}_{k \rightarrow l}^E}{\bar{K}_{k \rightleftharpoons l}^E} \, \bar{n}_{l} \, n_\mathrm{N} \right\rbrace \\
 & + \sum_{\substack{l \in \mathcal{K}_{\mathrm{BP}}\\ \left( k > l \right)}} \left\lbrace - \frac{\bar{k}_{l \rightarrow k}^E}{\bar{K}_{l \rightleftharpoons k}^E} \, \bar{n}_k \, n_\mathrm{N} + \bar{k}_{l \rightarrow k}^E \, \bar{n}_{l} \, n_\mathrm{N} \right\rbrace \\
 & -\bar{k}_k^{Df} \, \bar{n}_k \, n_\mathrm{N} + \frac{\bar{k}_k^{Df}}{\bar{K}_k^D} \, n_\mathrm{N}^3, \qquad k \in \mathcal{K}_{\mathrm{BP}}, \label{eq:molecular_nitrogen_bins_source_term}
 \end{split}
\end{equation}
where the first two rows on the right hand side of Eq.~(\ref{eq:molecular_nitrogen_bins_source_term}) account for production/depletion of $\mathrm{N_2}\left(k\right)$ due to excitation/deexcitation reactions of the type $\mathrm{N_2} \left( k \right) + \mathrm{N} \rightleftharpoons \mathrm{N_2} \left( l \right) + \mathrm{N}$, while the third accounts for production/depletion of $\mathrm{N_2}\left(k\right)$ due to dissociation/recombination reactions of the type $\mathrm{N_2} \left( k \right) + \mathrm{N} \rightleftharpoons 3 \mathrm{N}$. The bin-specific rate coefficients for excitation $\bar{k}_{k \rightarrow l}^E$, dissociation $\bar{k}_k^{Df}$, as well as the equilibrium constants relating bin populations were defined in Sec.~IIIA of our companion paper\cite{torres18a}.

The corresponding mass production term for atomic nitrogen, due to the same set of dissociation/recombination reactions, is given by:
\begin{equation}
 \frac{\omega_\mathrm{N}}{m_{\mathrm{N}}} = 2 \, \sum\limits_{k \in \mathcal{K}_{\mathrm{BP}}} \left\lbrace \bar{k}_{k}^{Df} \, \bar{n}_{k} \, n_{\mathrm{N}} - \frac{\bar{k}_{k}^{Df}}{\bar{K}_{k}^{D}} \, n_{\mathrm{N}}^3 \right\rbrace. \label{eq:atomic_nitrogen_reactive_source_term_bins}
\end{equation}

\section{Definition of internal temperature in thermal nonequilibrium} \label{sec:internal_temperature_definition}

Along with the simulations presented in this paper we report an ``internal'' temperature as a rough measure of the rovibrational energy content of $\mathrm{N_2}$. Note that this temperature is obtained \emph{a posteriori} from the nonequilibrium rovibrational bin populations, which emerge naturally in both the master equation calculations in Sec.~\ref{sec:parametric_study_kinetics} and the DSMC simulations in Sec.~\ref{sec:dsmc_verification_test_case}. Thus, $T_\mathrm{int}$ is never a direct input to the simulations themselves. Furthermore, the definition discussed here should only be applied in its current form to a gas containing a single diatomic species with a single internal energy mode of discrete rovibrational bins.

The thermal energy per unit mass contributed by the internal mode of $\mathrm{N_2}$ for an arbitrary distribution of uniform rovibrational bins is given by:
\begin{equation}
 e_\mathrm{N_2}^\mathrm{int} = \frac{1}{m_\mathrm{N_2}} \sum\limits_{k \in \mathcal{K}_\mathrm{BP}} \frac{\bar{n}_k}{n_\mathrm{N_2}} \bar{E}_k \label{eq:n2_average_internal_energy_bins}
\end{equation}

We now seek the hypothetical Boltzmann distribution of bin populations at the unknown $T_\mathrm{int}$, which would produce the same $e_\mathrm{N_2}^\mathrm{int}$ as the actual nonequilibrium distribution of Eq.~(\ref{eq:n2_average_internal_energy_bins}). These hypothetical populations would be given by:
\begin{equation}
 \left[ \frac{\bar{n}_k}{n_\mathrm{N_2}} \right]_{T_\mathrm{int}} = \frac{\bar{g}_k}{\bar{Q}_\mathrm{N_2}^\mathrm{int} \left( T_\mathrm{int} \right)} \exp \left( \frac{- \bar{E}_k}{\mathrm{k_B} \, T_\mathrm{int}} \right), \, k \in \mathcal{K}_\mathrm{BP}, \label{eq:boltzmann_distribution_at_internal_temperature}
\end{equation}
where the partition function of the internal energy mode is given by: 
\begin{equation}
 \bar{Q}_\mathrm{N_2}^\mathrm{int} \left( T_\mathrm{int} \right) = \sum\limits_{k \in \mathcal{K}_\mathrm{BP}} \bar{g}_k \, \exp \left( \frac{- \bar{E}_k}{\mathrm{k_B} \, T_\mathrm{int}} \right) \label{eq:bin_internal_mode_partition_function}
\end{equation}

Writing down the specific internal energy for this Boltzmann distribution yields:
\begin{equation}
 e_\mathrm{N_2}^\mathrm{int} \left( T_\mathrm{int} \right) = \frac{1}{m_\mathrm{N_2}} \sum_{k \in \mathcal{K}_\mathrm{BP}} \frac{\bar{g}_k \, \bar{E}_k}{\bar{Q}_\mathrm{N_2}^\mathrm{int} \left( T_\mathrm{int} \right)} \exp \left( \frac{- \bar{E}_k}{\mathrm{k_B} \, T_\mathrm{int}} \right) \label{eq:bin_internal_mode_thermal_energy_boltzmann}
\end{equation}
The actual nonequilibrium populations $\bar{n}_k / n_\mathrm{N_2}$ do not conform to the Boltzmann distribution of Eq.~(\ref{eq:boltzmann_distribution_at_internal_temperature}), and in this general case the thermal energy of $\mathrm{N_2}$ has to be calculated from Eq.~(\ref{eq:n2_average_internal_energy_bins}). However, we now equate Eq.~(\ref{eq:bin_internal_mode_thermal_energy_boltzmann}) with Eq.~(\ref{eq:n2_average_internal_energy_bins}) to obtain:
\begin{equation}
 \begin{split}
  \sum\limits_{k \in \mathcal{K}_\mathrm{BP}} & \left\lbrace \frac{\bar{n}_k}{n_\mathrm{N_2}} \, \bar{E}_k \right\rbrace \\
  & = \sum\limits_{k \in \mathcal{K}_\mathrm{BP}} \left\lbrace \frac{\bar{g}_k \, \bar{E}_k}{\bar{Q}_\mathrm{N_2}^\mathrm{int} \left( T_\mathrm{int} \right)} \exp \left( \frac{- \bar{E}_k}{\mathrm{k_B} \, T_\mathrm{int}} \right) \right\rbrace \label{eq:internal_temperature_definition}
 \end{split}
\end{equation}
With the ratios $\bar{n}_k / n_\mathrm{N_2}$ on the left hand side given by the actual nonequilibrium populations in the gas, this produces an equation with $T_\mathrm{int}$ as the only unknown. By iteratively adjusting the value of $T_\mathrm{int}$ until Eq.~(\ref{eq:internal_temperature_definition}) is satisfied for the nonequilibrium bin populations $\bar{n}_k / n_\mathrm{N_2}$, the correct energy content of the internal mode is maintained. Thus, in the general nonequilibrium case, the internal temperature must be found as an implicit solution to this non-linear equation.



\bibliography{rgd}{}
\bibliographystyle{unsrt}

\end{document}